\documentclass[footinbib, aps,pre,twocolumn,showpacs,amsfonts,amssymb,floatfix, superscriptaddress, reprint]{revtex4-1}
\usepackage[T1]{fontenc}
\usepackage[utf8]{inputenc}
\usepackage{float}
\usepackage{graphicx}
\usepackage{amssymb}
\usepackage{amsmath}
\usepackage{enumerate}
\usepackage{totcount}
\usepackage{float}

\makeatletter

\newcommand*{\diff}{\mathop{\!\mathrm{d}\!}}

\newcount\hour \newcount\minute
\hour=\time \divide \hour by 60 \minute=\time \count99=\hour
\multiply \count99 by -60 \advance\minute by \count99 \hour=\time
\divide \hour by 60
\date{October 9, 2019} 

\makeatother

\begin{document}

\title{Multiple Particle Correlation Analysis of Many-Particle Systems:\\
Formalism and Application to Active Matter}

\author{Rüdiger Kürsten}
\affiliation{Institut für Physik, Universität Greifswald, Felix-Hausdorff-Str. 6, 17489 Greifswald, Germany}
\author{Sven Stroteich}
\affiliation{Institut für Physik, Universität Greifswald, Felix-Hausdorff-Str. 6, 17489 Greifswald, Germany}
\author{Martín Zumaya Hérnandez}
\affiliation{Instituto de Ciencias Físicas, Universidad Nacional Autónoma de México, \\ Apartado Postal 48-3, Código Postal 62251, Cuernavaca, Morelos, México}
\affiliation{Centro de Ciencias de la Complejidad, Universidad Nacional Autónoma de México, Ciudad de México, México}
\author{Thomas Ihle}
\affiliation{Institut für Physik, Universität Greifswald, Felix-Hausdorff-Str. 6, 17489 Greifswald, Germany}


\begin{abstract}
	We introduce a fast spatial point pattern analysis technique which is suitable for systems of many identical particles giving rise to multi-particle correlations up to arbitrary order. The obtained correlation parameters allow to quantify the quality of mean field assumptions or theories that incorporate correlations of limited order. We study the Vicsek model of self-propelled particles and create a correlation map marking the required correlation order for each point in phase space incorporating up to ten-particle correlations. We find that multi-particle correlations are important even in a large part of the disordered phase. Furthermore, the two-particle correlation parameter serves as an excellent order parameter to locate both phase transitions of the system, whereas two different order parameters were required before.
\end{abstract}
\maketitle

In statistical physics, many-particle systems can be studied numerically using Molecular Dynamics or Monte Carlo simulations.
On the analytical side only a few examples such as the two-dimensional Ising model \cite{Onsager44} are known to be solvable exactly.
Therefore most analytical calculations are based on approximations.
Mean field approximations are among the most widely applied approaches both in- and out of equilibrium. 
The main idea is to replace the interactions of a given particle with all others by a single interaction with a mean field.

To improve mean field theories, low order correlations between two or more particles can be taken into account such as in the ring-kinetic theory of Ref.~\cite{CI15}. Other authors considered also third and higher order correlations, for example \cite{DA89, BCS89}. 
An alternative approach is the Minkowski functional \cite{MBW94}.
It is a technique in the analysis of spatial point patterns \cite{Diggle83, IPSS08, BRT15}, which has applications in diverse fields such as ecology \cite{LIBGGG09, VMGMW16}, neuroanatomy \cite{DLB91}, astronomy \cite{KSRBBGMPW97}, cell biology \cite{PSRMSL13}, image analysis \cite{MJM08}, etc. 

In real systems, all orders of correlations build up due to interactions. 
Nevertheless the wide use of mean field theories is motivated twofold. 
On the one hand, in many cases one is satisfied with a qualitative description of the system.
On the other hand, correlations might be either very small or not that important for the dynamics of the mean field such that neglecting them creates only a small error.

One aim of this Letter is to provide an efficient technique for experimental and simulation data to quantify the validity of assumptions on multi-particle correlations, not only two- and three-particle correlations but up to arbitrary order.
In principle, one could measure the full correlation functions, but this has several disadvantages.
First of all, correlation functions are high dimensional. 
Even two- or three-particle correlations can already depend on many degrees of freedom such as the positions and momenta of the involved particles.
Hence it is computationally expensive to measure them with high precision.
Secondly, assuming the full correlation function is measured accurately, it is not at all clear how important it is for the dynamics of certain observables.

In many cases, there is an intrinsic length scale $R$ associated with the interactions between particles, for example the Debye-length in plasma.
It is intuitive to study correlations on that length scale because only the nearby particles directly influence the dynamics of a given particle.
Hence, it is a minimum requirement for an accurate theory to reproduce the correct number of neighbors that reside within distance $R$ around a given particle. Clearly, this neighbor number is a random variable. We call its probability measure \textit{neighbor distribution} (ND).

In this Letter, we extract the minimal amount of information from the multi-particle correlation functions that is necessary to reproduce the correct ND.
We find that each correlation order contributes exactly two parameters to the parametrization of the ND.
Efficient techniques are developed to sample these parameters.
We derive the \textit{exact} ND by taking into account multi-particle correlations of finite but arbitrary order. Comparing the calculated distribution to the measured one, we obtain a lower bound on the correlation order that is required for an accurate description of the system.

We apply the analysis to a two-dimensional model of aligning self-propelled particles, the Vicsek model \cite{VCBCS95, CSV97} which is known to exhibit two different phase transitions \cite{GC04, CGGR08,SCT15, SCBCT15}. 
We measure a quantitative \textit{Correlation Map} of the model by marking the minimal required correlation order for each point in phase space. It is observed that even in the disordered phase multi-particle correlations are important in a large parameter range and the mean field assumption is valid only for very large noise strengths.
In addition, we find that a two-particle correlation parameter serves as an excellent order parameter to investigate both phase transitions whereas in the past two different order parameters have been used \cite{SCT15}.

In the following we assume that the system is spatially homogeneous and consists of many ($N \rightarrow \infty$) identical particles.
First, we ask for the number of particles within distance $R$ around an arbitrary fixed point in space which is in general not coinciding with any particle position.
If the particles are statistically independent, this number is Poisson-distributed \cite{Poisson1837} with mean $C_1=B_d(R) \rho$, where $\rho=N/V$ is the density and $B_d(R)$ the volume of a ball of radius $R$ in $d$ dimensions.

Relaxing the assumption of statistical independence and taking two-particle but no higher order correlations into account, we find the probability of $s$ particles to be within distance $R$ to an arbitrary fixed point as \footnote{\label{fn:supp}See supplemental material at pages 7-19}
\begin{align}
	p(s)=& 
	(C_1-C_2)^s \exp(C_2/2-C_1)
	\notag
	\\
	&\times \sum_{k=0}^{\infty} \bigg[ \frac{C_2}{2(C_1-C_2)^2}\bigg]^{k}
	 \frac{1}{k!(s-2k)!},
	\label{eq:poissonlike}
\end{align}
where we use the convention that $1/l!=0$ for $l<0$. 
Except for the mean number of particles in a ball, $C_1$, this distribution depends also on the correlation parameter $C_2$ where $C_k$ in general is defined by
\begin{align}
	&C_k:= N^k \int_{}^{} G_k(1,...,k)\prod_{l=1}^k \theta(R-|\mathbf{r}_l|)\diff l,
	\label{eq:corrcircle}
\end{align}
and where $\theta$ represents the Heaviside function, $G_1:=P_1$ and $G_k$ represents the $k$-particle correlation function (cumulants of it are Ursell-functions \cite{Ursell27}) for $k\ge2$. For example, the two-particle correlation function is defined as
\begin{align}
	 G_2(1,2):= P_2(1,2)-P_1(1)P_1(2),
	\label{eq:twoparticlecorrelation}
\end{align}
which is an example of the so-called cluster-expansion \cite{MM41}. Here, the argument ``$1$'' represents all degrees of freedom of particle $1$ and so on, and $P_1$ and $P_2$ are the one- and two-particle probability density functions, respectively.
It should be noted that the sum in Eq.~\eqref{eq:poissonlike} can also be expressed in terms of special functions.
We explicitly calculate the characteristic function of the probability distribution \eqref{eq:poissonlike} and generalize it by incorporating not only two-particle correlations but correlations up to an arbitrary but finite order $l_{max}$ to obtain \footnotemark[1]
\begin{align}
    \chi(u)= \exp\bigg[ \sum_{l=1}^{l_{max}} \sum_{t=0}^l (-1)^{l+t} \frac{C_l}{l!} \binom{l}{t} \exp(itu) \bigg],
    \label{eq:characteristicfunctiongeneral}
\end{align}{}
Note that $C_2=C_3=\dots =C_{l_{max}}=0$ yields the characteristic function of a Poisson distribution (PD) with mean $C_1$, which we recover also by setting $C_2=0$ directly in Eq.~\eqref{eq:poissonlike}.
Hence we call $\chi(u)$ from Eq.~\eqref{eq:characteristicfunctiongeneral} the charcteristic function of the correlation-induced \textit{generalized Poisson distribution}.
To the best of our knowledge, the distribution defined by~\eqref{eq:characteristicfunctiongeneral} has not been reported in the literature.

In the mean field scenario the \textit{neighbor distribution} (ND) equals the distribution of the number of particles within an arbitrarily located ball.
However, in the correlated case these distributions are different, and the characteristic function of the ND, the probability of an arbitrary particle to have $k$ neighbors $p_n(k)$, is related to $\chi(u)$ given in Eq.~\eqref{eq:characteristicfunctiongeneral} via
\begin{align}
    \chi_n(u)= \sum_{l=1}^{l_{max}} \sum_{t=0}^{l-1} &\binom{l-1}{t}(-1)^{t+l+1}\frac{1}{(l-1)!}
    \notag
    \\
    & \times D_l \exp(itu) \chi(u).
    \label{eq:characteristic_function_neighbors}
\end{align}
 We give a detailed derivation of Eqs.~\eqref{eq:characteristicfunctiongeneral} and \eqref{eq:characteristic_function_neighbors} in \footnotemark[1].
Here, the correlation coefficients $D_k$ are defined by
\begin{align}
D_k:= N^{k-1}\int G_k(1,2,\dots, k)\diff 1 \prod_{l=2}^k\theta(R-|\mathbf{r}_1-\mathbf{r}_l|)\diff l,
\label{eq:definitionDcoefficients}
\end{align}
where $D_1:=1$. Thus the correlation coefficients $C_k$ and $D_k$ are the desired minimal set of numbers necessary to calculate the correct ND.

We demonstrate how to extract these parameters from experimental or simulation data. 
In fact, they depend on two types of directly measurable quantities\footnotemark[1].
The first quantity $V_o(l)$ is the overlap volume of balls of radius $R$ drawn around each particle of an $l$-plet, summed over all ordered $l$-plets and normalized by the total system volume, see \footnotemark[1] for details. 
The second quantity $\mu_l$ is obtained as the expectation value of $s!/(s-l)!$ where $s$ is the number of neighbors of a randomly selected particle. 
Hence, $\mu_l$ can be easily sampled by counting the number of neighbors for each particle. 
The overlap can be either sampled by directly calculating the overlap volume of all $l$-plets or via an effective Monte Carlo algorithm that is described in \footnotemark[1] and does not depend on the spatial dimension. For the direct calculation in 2D we derive formulas for the overlap area of two and three \footnote{The overlap area of three arbitrarily placed circles generalizes the area formula of the famous Reuleaux triangle \cite{Reuleaux1875}.} circles depending only on the distances between the centers of the circles, for details see \footnotemark[1].
From those sampled quantities the relevant correlation parameters are obtained as \footnotemark[1]
\begin{align}
    C_l&= V_o(l) - \sum_{k=1}^{l-1} C_k V_o(l-k)\binom{l-1}{k-1},
    \label{eq:samplingCcoefficients}\\
    D_l &= \mu_{l-1} - \sum_{k=1}^{l-1} D_k V_o(l-k)\binom{l-1}{k-1}.
    \label{eq:samplingDcoefficients}
\end{align}
Remarkably, the above coefficients depend only on the correlation function $G_l$ and are not affected by modifications of lower or higher order correlations. 

The above theory is a powerful tool to systematically analyze multi-particle correlations of arbitrary order on the relevant length scale in many-particle systems.
Once the correlation coefficients $C_k$ and $D_k$ have been sampled, the ND can be calculated taking into account correlations up to the desired order.
Then, one can compare this calculated distribution to the directly measured one. 
Very good agreement would indicate that higher correlations might be neglected. 
However, if the calculated and measured ND totally disagree it is certain that higher order correlations are profoundly affecting the dynamics of the system and hence also its steady state.

We apply the correlation analysis to one of the prototypes of active matter, the standard Vicsek model in two dimensions, see \footnotemark[1] for the definition of the model.
Additionally, we apply the analysis technique to a continuous time variant of the Vicsek model that is closely related to direct experimental applications \cite{BCDDB13}. The results of this second model are qualitatively equivalent to the ones of the Vicsek model and are shown in \footnotemark[1].

\begin{figure}
	\includegraphics[width=0.9\columnwidth]{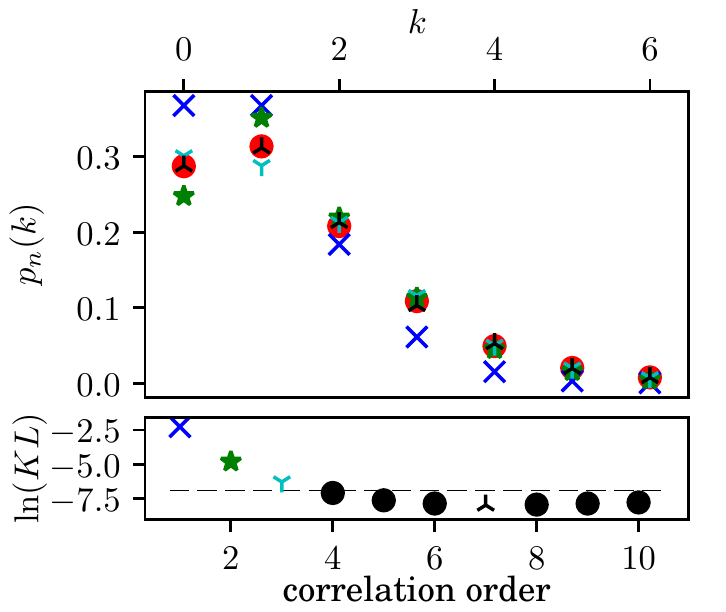}
	\caption{\textit{Top:} \textit{Neighbor distribution} (ND) for a randomly picked particle in the standard Vicsek model. Red circles display the measured distribution, blue crosses the Poisson distribution, green stars, cyan Y-shaped markers and black mirrored Y-shaped markers represent the distribution given by Eqs. \eqref{eq:characteristicfunctiongeneral}, \eqref{eq:characteristic_function_neighbors} including up to two-particle, up to three-particle and up to seven-particle correlations, respectively. System parameters: average neighbor number $C_1=1$, noise strength $\eta=0.48$, free path $v_0\tau=1$, interaction radius $R=1$, particle number $N=22500$. Averages have been calculated over $24$ realizations and $10^6$ time steps after a thermalization period of $10^5$ time steps for each realization. \textit{Bottom:} Logarithm of the Kullback-Leibler divergence between measured ND and calculated one using correlations up to different orders. \label{fig:neighbor_distribution_example}}
\end{figure}

In Fig.\ref{fig:neighbor_distribution_example} we present an example of the ND from a Molecular Dynamics simulation.
The red circles represent the directly measured distribution. 
The other symbols show distributions that have been calculated according to Eq.~\eqref{eq:characteristic_function_neighbors} taking into account various correlation orders.
We see that the Poisson distribution as well as the distribution that incorporates two-particle correlations do not agree very well with the measured distribution.
By additionally taking into account three-particle correlations the distribution is already quite close to the measured ND, and with up to seven-particle correlations it agrees almost perfectly.
We use the Kullback-Leibler divergence (KL) \cite{KL51} to quantify the agreement between measured and calculated distribution.
The logarithm of the KL is shown in the lower panel of Fig.~\ref{fig:neighbor_distribution_example} when different correlations orders are considered.
The agreement improves gradually until seven-particle correlations are considered.
For higher order correlations we obtain no improvement which can be understood looking at the accuracy with which the correlation parameters have been measured.
In Table~\ref{tab:table_example} we display the values of the correlation parameters with standard deviation of the mean in brackets for the same simulation as shown in Fig.~\ref{fig:neighbor_distribution_example}.
Up to seventh order the standard deviation is less then $1\%$ whereas it significantly increases for the next orders.
In principle, it is possible to measure even higher order correlation parameters accurately by analyzing significantly more data.
\begin{table}
\begin{tabular}{c|l|l}
     Correlation order $k$ & $\,\,\,\,C_k$ & $\,\,\,\,D_k$  \\
     \hline
     1 & 1 & 1 \\
     2 &0.3502115(92) & 0.4344897(60) \\
     3 & 0.230994(21) & 0.297475(21) \\
     4 & 0.236633(53) & 0.307351(80) \\
     5 & 0.34441(19) & 0.44878(31) \\
     6 & 0.6655(12) & 0.8678(16) \\
     7 & 1.616(11) & 2.102(10) \\
     8 & 4.70(12) & 6.121(91) \\
     9 & 15.7(1.1) & 20.69(86) \\
     10 & 59(10) & 78.0(8.1)
\end{tabular}
\caption{Correlation parameters measured in the same simulation as the one displayed in Fig.\ref{fig:neighbor_distribution_example} with standard deviation in brackets.\label{tab:table_example}}
\end{table}

\begin{figure}
	\includegraphics[width=0.9\columnwidth]{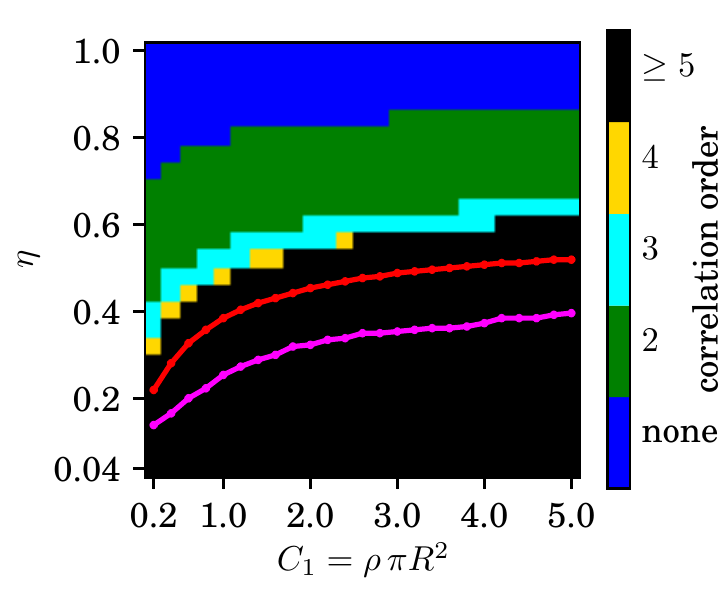}
	\caption{\textit{Correlation Map} providing the minimal order of correlations necessary to reproduce the measured \textit{neighbor distribution} (ND) up to a Kullback-Leibler divergence of $10^{-3}$ for the standard Vicsek model depending on noise strength $\eta$ and particle density. The transitions from disorder to polar ordered waves and from waves to a polar ordered homogeneous state are marked by the red and magenta line, respectively. Parameters and simulation time for each point as in Fig.~\ref{fig:neighbor_distribution_example}.
	\label{fig:corr_order}}
\end{figure}

To quantify the minimum order of correlations required to accurately reproduce the correct ND we set a threshold for the KL of $10^{-3}$ which is also displayed in the bottom part of Fig.~\ref{fig:neighbor_distribution_example} as dashed line. 
If the KL is less than the threshold we consider the agreement as good and if it is larger than the threshold we consider the agreement to be unsatisfying.
Using this definition we can determine the minimal order of correlations that need to be considered.
The results of extensive simulations for different values of noise strength and particle density are shown in Fig.~\ref{fig:corr_order} presenting a quantitative \textit{Correlation Map} of the standard Vicsek model.
We observe that a mean field hypothesis, as it is usually assumed when deriving field theories like e.g. \cite{TT95, TT98}, is valid only for very large noise strengths.
For smaller noise we need two-, three-, four-particle or even higher correlations.
Certainly, the required correlation order depends on the arbitrarily chosen threshold for the KL.
However, reasonable modifications of the threshold, e.g. $10^{-2}$ or $10^{-4}$, do not change the overall picture, see \footnotemark[1].

\begin{figure}
	\includegraphics[width=0.9\columnwidth]{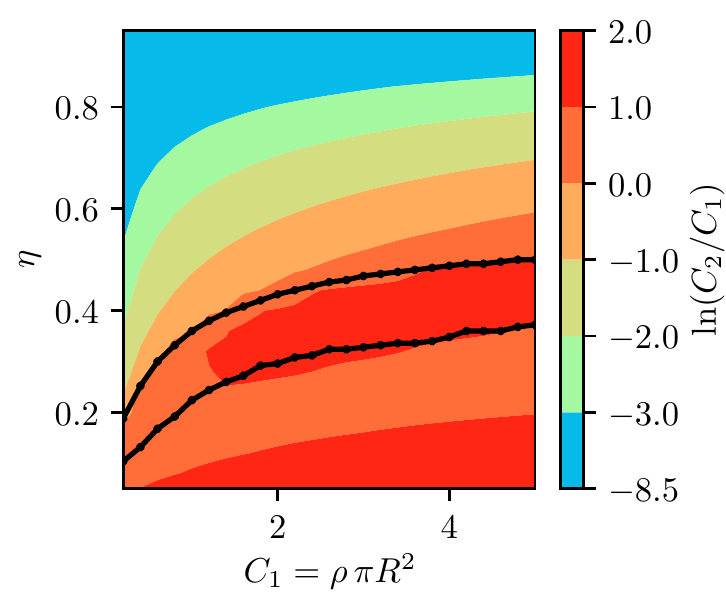}
	\caption{Logarithm of the ratio between two-particle correlation parameter $C_2$ and mean number of neighbors $C_1$ for the standard Vicsek model. The solid black lines are the same transition lines as in Fig.~\ref{fig:corr_order}. Data are from the same simulations.  \label{fig:c2}}
\end{figure}

\begin{figure}
	\includegraphics[width=0.9\columnwidth]{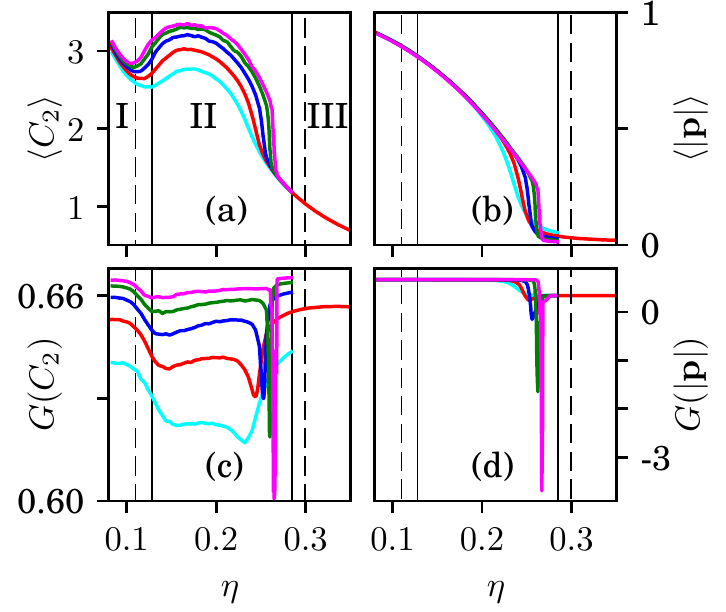}\\
		\includegraphics[width=0.9\columnwidth]{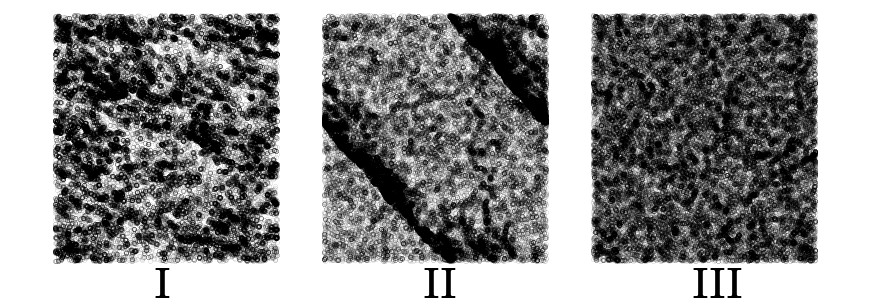}
\caption{\textit{Left:} average value (a) and Binder cumulant (c) of the two-particle correlation parameter $C_2$. \textit{Right:} average value (b) and Binder cumulant (d) of the polar order parameter. \textit{Bottom:} Snapshots in the homogeneous polar ordered (I), the polar ordered wave-front- (II) and the disordered (III) phase. The vertical dashed line between phases I and II represents the transition point from \cite{SCT15} for $N=10^4$ and the dashed line between II and III is the infinite system size transition of \cite{SCT15}. The solid vertical lines are the corresponding transitions obtained as the minima of the Binder cumulant of $C_2$ for the largest system $N=8\times 10^4$ (between I and II) and as an extrapolation to infinite system size (between II and III). Parameters: free path $v \tau=0.5$, density $C_1=\pi\rho=\pi/4$, particle number $N=5\times 10^3$ (cyan), $N=10^4$ (red), $N=2\times 10^4$ (blue), $N=4\times 10^4$ (green) and $N=8\times 10^4$ (magenta). Average over $24$ realizations, each started from random initial conditions, thermalized $10^5$ and recorded for $10^6$ time steps.  \label{fig:transitions}}
\end{figure}

We also investigate the strength of the correlation parameters.
As an example we present the two-particle correlation parameter $C_2$ as a function of density and noise strength in Fig.~\ref{fig:c2} and as a function of noise strength for fixed density in Fig.~\ref{fig:transitions} (a).
Surprisingly, $C_2$ is not a monotonous function of the noise strength for fixed particle density.
Instead it increases for decreasing noise strength until it reaches a maximum, goes through a minimum and then increases again.
This behavior can be understood by studying the phase transitions of the Vicsek model.
It is well known that there is  a transition from disordered motion at high noise to polar ordered collective motion at smaller noise.
The transition occurs discontinuously, has strong finite size effects and is closely related to the appearance of steep wave fronts \cite{CGGR08, Ihle13}.
It is absolutely plausible that these wave fronts go along with high correlations of many particles.

A second transition from the wave-like pattern to the formation of homogeneously distributed clusters, also called polar liquid or Toner-Tu-phase, occurs for even smaller noise strength \cite{GC04, CGGR08,SCT15}. This second transition goes along with a drastic drop of correlations, cf. Fig.~\ref{fig:transitions} (a) or Fig.~\ref{fig:c2}.
Thus we understand the remarkable high correlation island between the two transition lines in Fig.~\ref{fig:c2}.
Consequently, it is possible to use the correlation parameter $C_2$ as a single order parameter to study both transitions. In fact, we find two local minima in the Binder cumulant of the two-particle correlation parameter $C_2$, indicating the location of both transitions, cf. Fig.~\ref{fig:transitions} (c). Remarkably, this second transition is not at all observed when only the polar order parameter or its Binder cumulant is studied, see Fig.\ref{fig:transitions} (b) and (d).

In Fig.~\ref{fig:transitions} we chose a parameter set that was also studied in \cite{SCT15} for quantitative comparisons \footnote{In \cite{SCT15} the density was varied at constant noise strength.
Therefore both transitions occur not exactly at the same density $\rho=0.25$ that was used in Fig.~\ref{fig:transitions}.
The polar ordering transition occurred at $\rho=0.251$ for $\eta=0.3$ in Ref.~\cite{SCT15} from which we interpolated together with the next point $\eta_c=0.29944$ for $\rho=0.25$.}.
For other parameter sets, the two minima in the Binder cumulant are even much clearer, see \footnotemark[1].
The dashed horizontal lines of Fig.~\ref{fig:transitions} represent the transition points of Ref.~\cite{SCT15} whereas we obtained the solid vertical lines by studying the two-particle correlation order parameter.
The transition between polar liquid and wave-front phase was obtained in \cite{SCT15} as the end point of a hysteresis loop for a finite system ($N=10000$) which, however, depends on the details with which the hysteresis loop is recorded. It is therefore a different definition of the transition point than the minimum in the Binder cumulant that we used to obtain a slightly different value that is well-defined and independent on the measuring procedure.
The polar ordering transition was obtained in \cite{SCT15} for infinitely large systems $(\eta_c=0.29944)$ \footnotemark[3]. Therefore, we extrapolate our results to infinite systems by finite size scaling. We thus obtain $\eta_c=0.285$, however, more system sizes might be necessary to obtain an accurate \mbox{extrapolation \footnotemark[1]}.

Note that the computation time of the correlation analysis is of the order of the particle number times the average number of neighbors. This is the same complexity as for the Molecular Dynamics simulation itself. Hence, measuring the correlation parameters does not significantly increase the computation time.

In summary, we exactly calculate the neighbor distribution of a homogeneous many-particle system in the thermodynamic limit by including correlations of finite but arbitrary order and hence \textit{generalize} the Poisson distribution.
Explicit formulas that incorporate up to $k$-particles correlations are given.
The novel distributions depend only on a few correlation parameters.
We explicitly demonstrate how to sample these parameters by averaging numbers of neighbors and overlap volumes of balls.
The analysis is applied to the Vicsek model of self-propelled particles. 
We create a quantitative \textit{Correlation Map} of the model marking the minimal order of correlations required for each point in parameter space. 
Even in a large fraction of the disordered phase multi-particle correlations are important.
We find furthermore, that the two-particle correlation parameter serves as an order parameter to accurately investigate both phase transitions of the model.
We propose the use of the presented correlation analysis as a general technique in the study of phase transitions in and out of equilibrium and furthermore in the wide field of spatial point pattern analysis.

\begin{acknowledgments}
The authors gratefully acknowledge the GWK support for funding this project by providing computing time through the Center for Information Services and HPC (ZIH) at TU Dresden on the HRSK-II. The authors gratefully acknowledge the Universit\"atsrechenzentrum Greifswald and the HPC facilities at C3-UNAM for providing computing time. M.Z. acknowledges Francisco Sevilla and Maximino Aldana for valuable discussions and CONACyT scholarship 592409/307643. We thank Michael Himpel for valuable discussions. We thank Alexandre Solon for valuable discussions and providing data of Ref.\cite{SCT15}.
\end{acknowledgments}


%

\makeatletter

\setcounter{equation}{0}
\setcounter{figure}{0}
\renewcommand{\theequation}{S.\arabic{equation}}
\renewcommand{\thefigure}{S\arabic{figure}}

\clearpage
\makeatother

\title{Supplemental material to\\
Multiple Particle Correlation Analysis of Many-Particle Systems:\\
Formalism and Application to Active Matter}

\begin{center}
	\large{\textbf{Supplemental material to\\
Multiple Particle Correlation Analysis of Many-Particle Systems:\\
Formalism and Application to Active Matter}}
\end{center}

\section{Structure of the Supplemental}

This supplemental material is organized as follows.
In Sec.~\ref{sec:corr} we give the details of the derivation of the main theoretical result, Eqs.(4) and (5).
In Sec.~\ref{sec:sampling} we provide efficient algorithms to sample the parameters of the distribution given by Eqs.~(2) and (6).
In Sec.~\ref{sec:overlap} we derive explicit formulas to calculate the overlap area of two and three disks in two dimensions.
In Sec.~\ref{sec:models} we precisely define the models that are investigated in this paper and present additional numerical results. 

\section{Spatial Correlations\label{sec:corr}}

We assume that we have $N$ identical particles that are in a random state.
Furthermore we assume that $N$ is large such that is reasonable to consider the limit $N\rightarrow \infty$ when necessary.
We denote the $N$-particle probability density by $P_N(1,2, ..., N)$, where $1$ represents all degrees of freedom of particle one like e.g. position, velocity or orientation. 
Similarly $2, ..., N$ represent all degrees of freedom of particles $2$ to $N$.
We assume that $P_N$ is invariant under permutations of its arguments which should be the case at least in a stationary state.
The $k$-particle probability density is obtained from $P_N$ by
\begin{align}
    &P_k(1,2,...,k)
    \notag
    \\
    &:=\int P_N(1, 2, ..., N) \diff (k+1) \diff (k+2) \dots \diff N,
\end{align}{}
where the integration is performed over all possible configurations of the $N-k$ particles.
We assume furthermore that the one-particle probability density $P_1$ is spatially homogeneous, that is it does not depend on the position of the considered particle.
As long as there are no deterministic external forces and no walls all those assumption must be satisfied for finite ergodic systems in the stationary state.

We define the $k$-particle correlation function $G_k$ by
\begin{align}
    G_1(1) :=& P_1(1),
    \notag
    \\
    G_k(1,...,k):=&P_k(1,...,k)
    \notag
    \\
    &- \text{all possible combinations of}
\notag
\\
&\{G_1, ...,G_{k-1}\}, 
\notag
\\
&\text{where each of the arguments }\{1,..,k\} 
\notag
\\
&\text{appears exactly once}.
\label{eq:BBGKY}
\end{align}{}
For example we have for two-particle and three-particle correlation function
\begin{align}
    G_2(1,2):=&P_2(1,2)-P_1(1)P_1(2),
    \tag{$3$}
    \\
    G_3(1,2,3):=&P_3(1,2,3)-P_1(1)G_2(2,3)-P_1(2)G_2(1,3)
    \notag
    \\
    &-P_1(3)G_2(1,2)-P_1(1)P_1(2)P_1(3),
\end{align}
which is also called cluster-expansion.
On the other hand one can also obtain the $k$-particle probability from the one-particle probability and the correlation functions.
For example one finds the recursion relation
\begin{align}
    P_k(1,...,k)=&\sum_{l=1}^{k} G_l(1,...,l)P_{k-l}(l+1,...,k)
    \notag
    \\
    &+\text{Permutations},
    \label{eq:expansioncorrelationhierarchy}
\end{align}{}
where we sum over permutations of $\{2,..,k\}$ that are not only interchanging the arguments of each function, but exchange at least one argument of $G_l$ and $P_{k-l}$.

From the above definitions it is straight forward to find
\begin{align}
    &\int G_2(1,2)\diff 1= P_1(2)-P_1(2)=0
    \notag
    \\
    &=P_1(1)-P_1(1)=\int G_2(1,2)\diff 2.
\end{align}{}
Similarly, one can show by induction that in general
\begin{align}
	&\int_{}^{}G_k(1,\dots,k) \diff 1 = \int_{}^{}G_k(1,\dots,k) \diff 2 = \dots 
	\notag
	\\
	&=\int_{}^{}G_k(1,\dots,k) \diff N = 0.
	\label{eq:corrproperty}
\end{align}
The mean field assumption
\begin{align}
	P_N(1, 2, \dots, N)&= P_1(1)P_1(2) \dots P_1(N)
	\label{eq:meanfieldassumption}
\end{align}
is identical to setting all correlation functions to zero
\begin{align}
	G_2=G_3=\dots=G_N=0.
	\label{eq:meanfieldassumption2}
\end{align}
Considering the full hierarchy of correlation functions is extremely complex and the question arises which of the correlation functions are important and have to be considered.
Is mean field already an excellent approximation?
Or is it necessary to consider at least $G_2$?
Or also $G_3$?
Or do we need to take all correlations into account to obtain a good description of the system?

The correlation functions $G_2, G_3, \dots, G_N$ are high dimensional and therefore hard to measure.
Furthermore, it is not clear if it is important or not for the dynamics of $P_1$ if one of the correlation functions is large at some particular values of its arguments.

However, the microscopic dynamics of a particle depends explicitly on the number of its neighbors.
Hence, we know for sure that the probability distribution of the number of neighbors has an explicit influence on the dynamics of $P_1$.
Therefore we study the influence of the correlation functions $G_2, G_3, \dots$ on the number of neighbor probability distribution.

\subsection{Bernoulli experiment}
The mean field assumption \eqref{eq:meanfieldassumption2} corresponds to the random placement of independent particles. 
Asking for the number of particles in a circle of radius $R$ we will find the answer to be Poisson-distributed in the limit $N \rightarrow \infty$. 
In the mean field case the number of neighbors of a given particle have the same probability distribution.

Here, however, we take correlations into account.
As a first step we consider a nonzero $G_2$, however, we assume that higher correlations are zero, that is $G_3=G_4=\dots=G_N=0$.
In this case, we can still calculate the distribution of particles within a circle exactly in the limit of $N \rightarrow \infty$.
It is useful to define the parameter $C_k$ according to
\begin{align}
	&C_k:= N^k \int_{}^{} G_k(1,...,k)\prod_{l=1}^k \theta(R-|\mathbf{r}_l|)\diff l.
	\label{eq:corrcircle_main}
	\tag{$2$}
\end{align}
In particular we get for the two-particle correlations
\begin{align}
	&C_2:= N^2 \int_{}^{} G_2(1,2)\diff 1 \, \diff 2 \, \theta(R-|\mathbf{r}_1|)\theta(R-|\mathbf{r}_2|),
	\label{eq:corrcircle_supp}
\end{align}
where $\theta$ denotes the Heavyside function.
It is worth noting that changing one of the spatial integration regions to the outside of the circle yields 
\begin{align}
	-C_2&= N^2 \int_{}^{} G_2(1,2)\diff 1 \diff 2 \theta(R-|\mathbf{r}_1|)[1-\theta(R-|\mathbf{r}_2|)]
	\notag
	\\
	&= N^2 \int_{}^{} G_2(1,2)\diff 1 \diff 2 [1-\theta(R-|\mathbf{r}_1|)]\theta(R-|\mathbf{r}_2|)
	\label{eq:corrcircle2}
\end{align}
due to the property \eqref{eq:corrproperty} of $G_2$. Integrating both spatial coordinates outside the circle yields again
\begin{align}
	C_2= N^2 \int_{}^{} G_2(1,2)\diff 1 \diff 2 &[1-\theta(R-|\mathbf{r}_1|)]
	\notag
	\\
	&\times [1-\theta(R-|\mathbf{r}_2|)].
	\label{eq:corrcircle3}
\end{align}
The probability to find $s$ particles within a circle is given by
\begin{align}
	p(s) =& \binom{N}{s}\int \diff 1 \dots \diff N  P_N(1, \dots, N) \theta_{1} \dots \theta_{s} 
	\notag
	\\
	&\times(1-\theta_{s+1}) \dots (1-\theta_{N}),
	\label{eq:poissonlike1}
\end{align}
where we introduced the abbreviation
\begin{align}
    \theta_k:= \theta(R-|\mathbf{r}_k|)
    \label{eq:abbrtheta1}
\end{align}{}
and we included the combinatorial factor $\binom{N}{s}$ to take care of the fact that any selection of $s$ particles can be integrated over the inside of the circle and the rest over the outside. 
Inserting a sum over all possible combinations of $P_1$ and $G_2$ for $P_N$, that is setting $G_3=G_4=\dots=G_N=0$ in the full hierarchy of $P_N$, Eq.~\eqref{eq:BBGKY}, we obtain
\begin{align}
	&p(s)= \sum_{k_2=0}^{\infty} \bigg( \frac{C_2}{N^2} \bigg)^{k_2} \binom{N}{2k_2} \binom{2k_2}{k_2} \frac{k_2!}{2^{k_2}} 
	\notag
	\\
	&\times\sum_{k_1=0}^{\infty} \bigg(- \frac{C_2}{N^2} \bigg)^{k_1} \binom{N-2k_2}{2k_1} \binom{2k_1}{k_1} k_1!
	\notag
	\\
	&\times\sum_{k_0=0}^{\infty} \bigg(\frac{C_2}{N^2} \bigg)^{k_0} \binom{N-2k_2-2k_1}{2k_0} \binom{2k_0}{k_0} \frac{k_0!}{2^{k_0}}
	\notag
	\\
	&\times\sum_{q_1=0}^{\infty} \bigg(\frac{C_1}{N} \bigg)^{q_1} \binom{N-2k_2-2k_1-2k_0}{q_1} 
	\notag
	\\
	&\times\sum_{q_0=0}^{\infty} \bigg(1-\frac{C_1}{N} \bigg)^{q_2} \delta(s-2k_2-k_1-q_1) 
	\notag
	\\
	& \times \delta(N-2k_2-2k_1-2k_0-q_1-q_0) 
	\label{eq:poissonlike2}
\end{align}
with the convention that $\binom{N}{l}=0$ if $l>N$.
$k_2$ represents the number of pairs ($G_2$) where both arguments are integrated over the inside of the circle and the following combinatorial factor is the number of possibilities to choose $k_2$ pairs from $N$ particles.
$k_1$ is the number of pairs with one argument integrated over the circles inside and one argument integrated over the outside.
The following combinatorial factor gives the number of possible choices of $k_1$ ordered pairs from the remaining $N-2k_2$ particles.
$k_0$ represents the number of pairs where both arguments are integrated over the outside of the circle, analogously $q_1$ and $q_0$ are the numbers of single particles ($P_1$) integrated over the circles inside and outside, respectively. 
There, we also include the corresponding combinatorial factors. 
The first $\delta$-function takes care that there are exactly $s$ particles inside and the second one, that there are exactly $N$ particles in total.
All the combinatorial factors can be combined to
\begin{align}
	\frac{N!}{(N-2k_0-2k_1-2k_2-q_1)! q_1! k_0! k_1! k_2! 2^{k_0} 2^{k_2}}.
	\label{eq:poissonlike3}
\end{align}
For large $N$ we can approximate
\begin{align}
	&\frac{N!}{(N-2k_0-2k_1-2k_2-q_1)!}= N \cdot(N-1) \cdot \dots 
	\notag
	\\
	&\times(N-2k_0-2k_1-2k_2-q_1 +1 ) 
	\notag
	\\
	&\approx N^{2k_0+2k_1+2k_2+q_1}.
	\label{eq:poissonlike4}
\end{align}
All corrections are of lower order in $N$ and are eventually going to zero as $N \rightarrow \infty$. Inserting \eqref{eq:poissonlike4} in \eqref{eq:poissonlike2} and performing the sum over $q_0$ with the second $\delta$ we obtain
\begin{align}
	&p(s)= \sum_{k_2=0}^{\infty}  \bigg( \frac{C_2}{2}\bigg)^{k_2} 
	\sum_{k_1=0}^{\infty} (- C_2)^{k_1} 
	\sum_{k_0=0}^{\infty} \bigg(\frac{C_2}{2}\bigg)^{k_0}
	\sum_{q_1=0}^{\infty} C_1^{q_1}  
	\notag
	\\
	&\times \bigg(1-\frac{C_1}{N} \bigg)^{N-2k_0-2k_1-2k_2-q_1} \frac{1}{k_0! k_1! k_2! q_1!} 
	\notag
	\\
	&\times\delta(s-2k_2-k_1-q_1).
	\label{eq:poissonlike5}
\end{align}
In the limit $N \rightarrow \infty$ we obtain
\begin{align}
	\bigg(1-\frac{C_1}{N} \bigg)^{N-2k_0-2k_1-2k_2-q_1} \rightarrow \exp(-C_1)
	\label{eq:poissonlike6}
\end{align}
such that we find
\begin{align}
	&p(s)= \sum_{k_0=0}^{\infty} \bigg(\frac{C_2}{2}\bigg)^{k_0}\frac{\exp(-C_2/2)}{k_0!} \sum_{k_2=0}^{\infty}  \bigg( \frac{C_2}{2}\bigg)^{k_2} 
	\notag
	\\
	&\times \sum_{k_1=0}^{\infty} (- C_2)^{k_1} 
	\sum_{q_1=0}^{\infty} C_1^{q_1}  
	\exp(-C_1) \exp(C_2/2) \frac{1}{k_1! k_2! q_1!} 
	\notag
	\\
	&\times\delta(s-2k_2-k_1-q_1).
	\label{eq:poissonlike7}
\end{align}
The sum over $k_0$ is just a sum over a Poisson distribution which gives one. Performing also the sum over $q_1$ with the $\delta$ we obtain
\begin{align}
	&p(s)= C_1^s \exp(-C_1) \exp(C_2/2)\sum_{k_2=0}^{\infty}  \bigg( \frac{C_2}{2 C_1^2}\bigg)^{k_2} \frac{1}{(k-2k_2)!}
	\notag
	\\
	&\times \frac{1}{k_2!}\sum_{k_1=0}^{k-2k_2} \bigg(- \frac{C_2}{C_1}\bigg)^{k_1} 1^{s-2k_2-k_1} 
	\frac{(s-2k_2)!}{k_1! (s-2k_2-k_1)!}. 
	\label{eq:poissonlike8}
\end{align}
Using the binomial theorem to perform the sum over $k_1$ we obtain
\begin{align}
	&p(s)= C_1^s \exp(-C_1) \exp(C_2/2)\sum_{k_2=0}^{\infty}  \bigg( \frac{C_2}{2 C_1^2}\bigg)^{k_2} 
	\notag
	\\
	&\times \frac{1}{k_2!(s-2k_2)!} \bigg(1 - \frac{A}{M}\bigg)^{k-2k_2}
	\notag
	\\
	&=(C_1-C_2)^s \exp(C_2/2-C_1)
	\notag
	\\
	&\times \sum_{k_2=0}^{\infty} \bigg[ \frac{C_2}{2(C_1-C_2)^2}\bigg]^{k_2}
	\frac{1}{k_2!(s-2k_2)!}
	\label{eq:poissonlike9}
	\tag{$1$}
\end{align}
Eventually, this sum can be expressed as
\begin{align}
	&p(s)= \frac{C_1^s \exp(-C_1)}{s!} \exp(C_2/2)    \bigg(1 - \frac{C_2}{C_1}\bigg)^{s}
	\notag
	\\
	&\times \bigg( - \frac{(C_2-C_1)^2}{2C_2}\bigg)^{(1-s)/2} 
	\notag
	\\
	&\times U[(1-k)/2, 3/2, -(C_2-C_1)^2/(2C_2)], 
	\label{eq:poissonlike10}
\end{align}
where $U(., ., .)$ is the confluent hypergeometric function of the second kind.

Considering the form of $p(s)$ given by Eq.~\eqref{eq:poissonlike9} it is possible to calculate the characteristic function $\chi(u):=\sum_s p(s) \exp(ius)$ exactly by first performing the sum over $s$ and than the remaining sum over $k_2$.
The surprisingly simple result is given by
\begin{align}
    &\chi(u)=\langle \exp(ius) \rangle 
    \notag
    \\
    &= \exp\bigg[ -C_1+\frac{C_2}{2} +(C_1-C_2)\exp(iu)
    \notag
    \\
    &\phantom{= \exp\bigg[}+\frac{C_2}{2}\exp(i2u)\bigg].
    \label{eq:characteristicfunction}
\end{align}{}
 
Taking into account nonzero three-particle correlations, however, still assuming $G_4=G_5=...=G_N=0$ we find in complete analogy to Eq.\eqref{eq:poissonlike2} the probability of finding exactly $s$ particles within a given circle of radius $R$ as
\begin{align}
	&p(s)= \sum_{l_0=0}^{\infty} \bigg( \frac{-C_3}{N^3} \bigg)^{l_0} \binom{N}{3l_0} \binom{3l_0}{l_0} \binom{2l_0}{l_0}\frac{(l_0!)^2}{6^{l_0}}
	\notag
	\\
	&\times \sum_{l_1=0}^{\infty} \bigg( \frac{C_3}{N^3} \bigg)^{l_1} \binom{N-3l_0}{3l_1} \binom{3l_1}{l_1} \binom{2l_1}{l_1}\frac{(l_1!)^2}{2^{l_1}}
	\notag
	\\
	&\times \sum_{l_2=0}^{\infty} \bigg( \frac{-C_3}{N^3} \bigg)^{l_2} \binom{N-3l_0-3l_1}{3l_2} \binom{3l_2}{l_2} \binom{2l_2}{l_2}\frac{(l_2!)^2}{2^{l_2}}
	\notag
	\\
	&\times \sum_{l_3=0}^{\infty} \bigg( \frac{C_3}{N^3} \bigg)^{l_3} \binom{N-3l_0-3l_1-3l_2}{3l_3}
	\notag
	\\
	&\times \binom{3l_3}{l_3} \binom{2l_3}{l_3}\frac{(l_3!)^2}{6^{l_3}}
	\notag
	\\
	&\times\sum_{k_0=0}^{\infty} \bigg( \frac{C_2}{N^2} \bigg)^{k_0} \binom{N-3l}{2k_0} \binom{2k_0}{k_0} \frac{k_0!}{2^{k_0}} 
	\notag
	\\
	&\times\sum_{k_1=0}^{\infty} \bigg(- \frac{C_2}{N^2} \bigg)^{k_1} \binom{N-3l-2k_0}{2k_1} \binom{2k_1}{k_1} k_1!
	\notag
	\\
	&\times\sum_{k_2=0}^{\infty} \bigg(\frac{C_2}{N^2} \bigg)^{k_2} \binom{N-3l-2k_0-2k_1}{2k_2} \binom{2k_2}{k_2} \frac{k_2!}{2^{k_2}}
	\notag
	\\
	&\times\sum_{q_0=0}^{\infty} \bigg(1-\frac{C_1}{N} \bigg)^{q_0} \binom{N-3l-2k}{q_0} 
	\notag
	\\
	&\times\sum_{q_1=0}^{\infty} \bigg(\frac{C_1}{N} \bigg)^{q_1} \delta(s-l_1-2l_2-3l_3-k_1-2k_2-q_1) 
	\notag
	\\
	& \times \delta(N-3l-2k-q), 
	\label{eq:poissonlike13}
\end{align}
where $l=l_0+l_1+l_2+l_3$, $k=k_0+k_1+k_2$ and $q=q_0+q_1$. Here $l_i$ denotes the number of correlated triplets with exactly $i$ particles inside the circle. Analogously $k_i$ and $q_j$ are the numbers of correlated pairs or singlets with exactly $i$ and $j$ particles inside the circle, respectively.
Simplifying the combinatorial factors we obtain
\begin{align}
	&p(s)= \sum_{\dots} \bigg( \frac{-C_3}{6} \bigg)^{l_0}\bigg( \frac{C_3}{2} \bigg)^{l_1}\bigg( \frac{-C_3}{2} \bigg)^{l_2}\bigg( \frac{C_3}{6} \bigg)^{l_3}\bigg( \frac{C_2}{2} \bigg)^{k_0}
	\notag
	\\
	&\times (-C_2)^{k_1}\bigg( \frac{C_2}{2} \bigg)^{k_2}\bigg(1-\frac{C_1}{N} \bigg)^{q_0} C_1^{q_1}
	\notag
	\\
	&\times \frac{N!}{ (N-3l-2k-q_0)! N^{3l+2k+q_1}}
	\notag
	\\
	&\times \frac{\delta(s-l_1-2l_2-3l_3-k_1-2k_2-q_1)}{l_0!l_1!l_2!l_3!k_0!k_1!k_2!q_0!}
	\notag
	\\
	&\times \delta(N-3l-2k-q).
	\label{eq:poissonlike14}
\end{align}
Performing the sum over $q_0$ with the second $\delta$ we obtain
\begin{align}
	&p(s)= \sum_{\dots} \bigg( \frac{-C_3}{6} \bigg)^{l_0}\bigg( \frac{C_3}{2} \bigg)^{l_1}\bigg( \frac{-C_3}{2} \bigg)^{l_2}\bigg( \frac{C_3}{6} \bigg)^{l_3}\bigg( \frac{C_2}{2} \bigg)^{k_0}
	\notag
	\\
	&\times (-C_2)^{k_1}\bigg( \frac{C_2}{2} \bigg)^{k_2}\bigg(1-\frac{C_1}{N} \bigg)^{N-3l-2k-q_1} C_1^{q_1} 
	\notag
	\\
	&\times \frac{N!}{ (N-3l-2k-q_1)! N^{3l+2k+q_1}}
	\notag
	\\
	&\times \frac{\delta(s-l_1-2l_2-3l_3-k_1-2k_2-q_1)}{l_0!l_1!l_2!l_3!k_0!k_1!k_2!q_1!}.
	\label{eq:poissonlike15}
\end{align}
Performing the limit $N\rightarrow \infty$ we obtain
\begin{align}
	\bigg(1-\frac{C_1}{N} \bigg)^{N-3l-2k-q_1}  &\rightarrow \exp(-C_1) 
	\notag
	\\
	\frac{N!}{ (N-3l-2k-q_1)! N^{3l+2k+q_1}} & \rightarrow 1.
	\label{eq:poissonlike16}
\end{align}
Performing the sums over $l_0$ and $k_0$
\begin{align}
	&\sum_{l_0=0}^{\infty} \frac{  ( -C_3/6 )^{l_0}}{l_0!} =\exp(-C_3/6),
	\notag
	\\
	&\sum_{k_0=0}^{\infty} \frac{ ( C_2/2 )^{k_0}}{k_0!} =\exp(C_2/2)
	\label{eq:poissonlike17}
\end{align}
we obtain
\begin{align}
	&p(s)= \sum_{\dots} \bigg( \frac{C_3}{2} \bigg)^{l_1}\bigg( \frac{-C_3}{2} \bigg)^{l_2}\bigg( \frac{C_3}{6} \bigg)^{l_3}(-C_2)^{k_1}
	\notag
	\\
	&\times \bigg( \frac{A}{2} \bigg)^{k_2} C_1^{q_1}\exp(-C_1+C_2/2-C_3/6) 
	\notag
	\\
	&\times \frac{\delta(s-l_1-2l_2-3l_3-k_1-2k_2-q_1)}{l_1!l_2!l_3!k_1!k_2!q_1!}.
	\label{eq:poissonlike18}
\end{align}
Performing the sum over $q_1$ with the $\delta$ and the binomial sums over $l_1$ and $k_1$ we find
\begin{align}
	&p(s)= \sum_{l_2 l_3 k_2}\exp(-C_1+C_2/2-C_3/6) \bigg[C_1 + \frac{C_3}{2} -C_2 \bigg]^s 
	\notag
	\\
	&\times \bigg[ - \frac{2C_3}{ (2C_1+C_3-2C_2)^2}  \bigg]^{l_2} \bigg[ \frac{4C_3}{3} \frac{1}{ (2C_1+C_3-2C_2)^3}  \bigg]^{l_3}
	\notag
	\\
	&\times \bigg[  \frac{2C_2}{ (2C_1+C_3-2C_2)^2}  \bigg]^{k_2} 
	\notag
	\\
	&\times \frac{1}{l_2!l_3!k_2!(s-2l_2-3l_3-2k_2)!}.
	\label{eq:poissonlike19}
\end{align}
It is possible to represent one more sum in terms of a confluent hypergeometric function, however, we can also handle the three remaining sums numerically. 
Furthermore, although Eq.~\eqref{eq:poissonlike19} looks much more complicated than Eq.~\eqref{eq:poissonlike9} it is still possible to calculate the characteristic function exactly by first performing the sum over $s$ and then all the remaining sums over $k_2, l_2, l_3$ resulting in
\begin{align}
    \chi(u)= \exp\bigg[&-C_1+\frac{C_2}{2}-\frac{C_3}{6}+ \bigg(C_1-C_2+\frac{C_3}{2}\bigg)e^{iu} 
    \notag
    \\
    &+ \bigg(\frac{C_2}{2}-\frac{C_3}{2}\bigg)e^{i2u}+\frac{C_3}{6}e^{i3u}\bigg].
\end{align}{}
Including additionally correlations of order $k$ the procedure is completely analogous.
There appear $k+1$ more sums in the analog of Eq.~\eqref{eq:poissonlike13}.
Two of them can be calculated immediately, such that one remains with $k-1$ additional sums.
However, calculating the characteristic function one can perform all this sums exactly and it is possible to show by induction that 
\begin{align}
    \chi(u)= \exp\bigg[ \sum_{l=1}^{l_{max}} \sum_{t=0}^l (-1)^{l+t} \frac{C_l}{l!} \binom{l}{t} \exp(itu) \bigg],
    \label{eq:characteristicfunctiongeneral_supp}
    \tag{$4$}
\end{align}{}

Eventually, the distribution of the number of particles found in a ball of Radius $R$ is obtained by transforming back the characteristic function
\begin{align}
    p(s)= \lim_{n\rightarrow \infty} \frac{1}{n}\sum_{u=0}^{n-1}\exp(-ius)\chi(u).
\end{align}

\subsection{Number of neighbor distribution}
If the particles were independent the distribution \eqref{eq:poissonlike10} would already give the neighbor distribution of a given particle. 
With nonzero pair correlations, however, the neighbor distribution differs.
The neighborhood distribution is given by
\begin{align}
	&p_n(k) = \langle \delta(k- \sum_{j=2}^{N} \theta_{1j})\rangle
	\notag
	\\
	&= \int_{}^{} P_N (1, 2, \dots) \delta(k- \sum_{j=2}^{N} \theta_{1j})
	\notag
	\\
	&= \int_{}^{} P_1(1) P_{N-1}(2,3,\dots) \delta(k- \sum_{j=2}^{N} \theta_{1j})
	\notag
	\\
	&+ \int_{}^{} \sum_{l=2}^{N} G_2(1,l) P_{N-2}(2,\dots l-1, l+1, \dots)
	\notag
	\\
	&\times \delta(k- \sum_{j=2}^{N} \theta_{1j})
	\notag
	\\
	&= \int_{}^{} P_1(1) P_{N-1}(2,3,\dots) \delta(k- \sum_{j=2}^{N} \theta_{1j})
	\notag
	\\
	&+ \int_{}^{} (N-1) G_2(1,2) P_{N-2}(3, 4, \dots) \delta(k- \sum_{j=2}^{N} \theta_{1j}),
	\label{eq:poissonlike11}
\end{align}
where integration is performed over all coordinates.
Similar to Eq.~\eqref{eq:abbrtheta1} we introduced the abrreviation
\begin{align}
    \theta_{jk}:= \theta(R-|\mathbf{r}_k-\mathbf{r}_j|).
    \label{eq:abbrtheta2}
\end{align}{}
In the limit $N \rightarrow \infty$ we can replace $N-1$ by $N$. 
The first term in the above expression corresponds to distribution of an Bernoulli experiment, however, the second term gives a correction
\begin{align}
&	p_n(k) = p(k)
\notag
\\
	&+ \int_{}^{} N G_2(1,2) \theta_{12} P_{N-2}(3, 4, \dots) \delta(k- \sum_{j=2}^{N} \theta_{1j})
	\notag
	\\
	&+ \int_{}^{} N G_2(1,2) (1-\theta_{12}) P_{N-2}(3, 4, \dots) \delta(k- \sum_{j=2}^{N} \theta_{1j})
	\notag
	\\
	& = p(k) 
	\notag
	\\
	&+ \int_{}^{} N G_2(1,2) \theta_{12} P_{N-2}(3, 4, \dots) \delta(k-1- \sum_{j=3}^{N} \theta_{1j})
	\notag
	\\
	&+ \int_{}^{} N G_2(1,2) (1-\theta_{12}) P_{N-2}(3, 4, \dots) \delta(k- \sum_{j=3}^{N} \theta_{1j})
	\notag
	\\
	&= p(k) + \int_{}^{} N G_2(1,2) \theta_{12} P_{N-2}(3, 4, \dots) 
	\notag
	\\
	& \times [\delta(k-1- \sum_{j=3}^{N} \theta_{1j})- \delta(k- \sum_{j=3}^{N} \theta_{1j})] ,
	\notag
	\\
	&= p(k) + [p(k-1)-p(k)] N \int_{}^{}G_2(1,2) \theta_{12},
	\label{eq:poissonlike12}
\end{align}
where it should be noted that $p(k)=0$ for $k<0$.

For generalizing this result taking into account correlations up to order $l_{max}$ we introduce the abbreviations
\begin{align}
D_k:= N^{k-1}\int G_k(1,2,..., k)\theta_{12}... \theta_{1k}\diff 1 ... \diff k,
\label{eq:definitionDcoefficients_supp}
\tag{$6$}
\end{align}
where $D_1=1$.
In analogy to Eqs.~\eqref{eq:poissonlike11} and \eqref{eq:poissonlike12} we find in this case
\begin{align}
    p_n(k)= \sum_{l=1}^{l_{max}} \sum_{t=0}^{l-1} \binom{l-1}{t}(-1)^{t+l+1}\frac{1}{(l-1)!} p(k-t)D_l.
\end{align}{}
It is straight forward to calculate the characteristic function of the number of neighbor distribution
\begin{align}
    \chi_n(u)= \sum_{l=1}^{l_{max}} \sum_{t=0}^{l-1} &\binom{l-1}{t}(-1)^{t+l+1}\frac{1}{(l-1)!}
    \notag
    \\
    & \times D_l \exp(itu) \chi(u),
    \tag{$5$}
\end{align}
where $\chi(u)$ is given by Eq.~\eqref{eq:characteristicfunctiongeneral_supp}.
Hence we expressed the number of neighbor distribution in terms of the coefficients $C_l$ and $D_l$.
The expression is exact in the limit $N \rightarrow \infty$ if the higher order correlation functions $G_l \equiv 0$ vanish for $l> l_{max}$.

\section{Sampling parameters $C_l$ and $D_l$\label{sec:sampling}}

It is not possible to sample the parameters $C_l$ and $D_l$ directly.
However, they are related to two types of quantities that can be sampled with high efficiency.
The first quantities are expectation values of polynomials of the number of neighbors
\begin{align}
\mu_l := \bigg \langle \frac{k!}{(k-l)!} \bigg \rangle = \sum_{k=0}^{N-1} p_n(k)\frac{k!}{(k-l)!}
\label{eq:defmu}
\end{align}
for $l=1,2,...$ .
It is straight forward to sample these quantities by just counting the number of neighbors for each particle in each time step and averaging the corresponding polynomial of this number.

The second quantities are sums of the overlap volume of balls of radius $R$ drawn around all ordered $l$-plets of different particles divided by the systems Volume
\begin{align}
    V_o(l) := \frac{1}{V} \bigg \langle \sum_{k_1, ..., k_l, k_i\neq k_j} \mathcal{V}_{\text{overlap}}(k_1,...,k_l) \bigg \rangle,
    \label{eq:overlap}
\end{align}{}
where $\mathcal{V}_{\text{overlap}}(k_1, ..., k_l)$ denotes the volume of the overlap of $l$ balls of radius $R$ centered at the positions of the particles $k_1, ..., k_l$.
There are two efficient ways to sample these quantities. 
One can either take all $l$-plets and calculate the overlap volume exactly.
For example for pairs, the overlap only depends on the distance between the particles and for triplets it depends only on the three distances between the three particles.
In two dimensions we derive explicit formulas for the two and three particle overlap areas in Sec.~\ref{sec:overlap}.

There is an alternative Monte Carlo approach that is in particular suitable to sample also $V_o(l)$ for large $l$.
Assume we we chose an ordered triplet of particles, e.g. $(7, 3, 5)$ and want to estimate $\frac{\mathcal{V}_\text{overlap}(7,3,5)}{V}$.
We can place a virtual particle at a random position and count one if the virtual particle lies in the overlap of $(7,3,5)$ that is if it has a distance smaller than $R$ to each of the particles $7,3,5$.
Otherwise we count zero.
If we repeat the procedure $n$ times and divide the sum by $n$ we obtain an estimate of the above mentioned volume fraction.

In order to obtain an estimate for the sum in Eq.~\eqref{eq:overlap}, in principle, we need to repeat the procedure for each ordered triplet.
We can instead just sum over all unordered triplets and multiply by $l!$, where $l=3$ in the case of triplets.
Furthermore, we reach an extreme improvement in performance using the same randomly placed virtual particles for all triplets.
In that way, we just need to find out in the overlap of how many triplets of real particles does a randomly placed real particle lie?
This number depends only on the number of real neighbors of the virtual particle.
Assume the virtual particle has $m$ real neighbors, than it lies in the overlap of $\binom{m}{l}$ unordered or $\binom{m}{l} l!$ ordered triplets of real particles, where $l=3$ for triplets.
In general we find 
\begin{align}
    V_o(l)= \bigg \langle \binom{m}{l} l! \bigg \rangle,
\end{align}
where $m$ is the number of real neighbors of a randomly placed virtual particle and the expectation value is taken over many virtual particles and a time series of the positions of the real particles.
Usually it does not significantly slow down molecular dynamics simulations if we use as many virtual particles as we have real particles.
In Table~\ref{tab:my_label} we show an example of a measurement of $V_o(2)$. We compare the values obtained by directly measuring the overlap areas with the Monte Carlo approach. The measured values agree within error bars. As expected, the uncertainty in the Monte Carlo sampling is a bit larger (by a factor of $1.5$) but on the same order of magnitude.
On the other hand, the Monte Carlo approach has the advantage that one can sample $V_o(l)$ for all $l$ at the same time.
\begin{table}[]
    \centering
\begin{tabular}{c|c}
     & $ V_o(2) $   \\
     \hline
    Direct Measurement & 7.67(17) \\
    Monte Carlo Sampling & 7.50(28)
\end{tabular}
\caption{Comparison of the direct measurement and the Monte Carlo sampling of $V_0(2)$ for the standard Vicsek model. We give the standard deviation of the mean value in brackets. Both approaches gives similar results, except the standard deviation of the Monte Carlo sampling is approximately $50\%$ larger. System parameters: $N=22500, \eta = 0.5, v\tau/R=5, C_1 = 5$, thermalization for $10^5$ and measurement for $10^6$ time steps for 25 realizations. }
    \label{tab:my_label}
\end{table}

We can directly connect the correlation coefficients $C_l$ to the sampled overlap areas.
We rewrite their definition as
\begin{align}
    C_l&= N^l \int G_l \theta_1 \dots \theta_l \diff 1 \dots \diff l
    \notag
    \\
    &= N^l \int G_l \theta_{01} \dots \theta_{0l} \diff 1 \dots \diff l
    \notag
    \\
    &= \frac{N^l}{V} \int G_l \theta_{01} \dots \theta_{0l} \diff 0\diff 1 \dots \diff l,
    \label{eq:samplingcorrelationcoefficients}
\end{align}{}
where we invented a virtual particle $0$ in the second line. Due to translational invariance the integral does not depend on the position of this virtual particle. Hence we can also integrate the virtual particles position over the whole space and compensate for it by dividing by the total volume.
For large $N$ we have furthermore
\begin{align}
    &\frac{N^l}{V} \int P_l \theta_{01} \dots \theta_{0l} \diff 0 \diff 1 \dots \diff l
    \notag
    \\
    &=\frac{N^l}{V} \int P_N \theta_{01} \dots \theta_{0l} \diff 0 \diff 1 \dots \diff N
    \notag
    \\
    &\approx V_o(l),
    \label{eq:samplingcorrelationcoefficients2}
\end{align}{}
since the integral divided by $V$ gives the probability that a virtual particle is in the overlap of the particles $(1,\dots,l)$.
However, for large $N$ there are approximately $N^l$ possibilities to choose an ordered $l$-plet from $N$ particles.

Inserting the expansion \eqref{eq:expansioncorrelationhierarchy} for $G_l$ into Eq.~\eqref{eq:samplingcorrelationcoefficients} and using Eq.~\eqref{eq:samplingcorrelationcoefficients2} we obtain
\begin{align}
    C_l= V_o(l) - \sum_{k=1}^{l-1} C_k V_o(l-k)\binom{l-1}{k-1},
    \label{eq:samplingCcoefficients_supp}
    \tag{$7$}
\end{align}
where $\binom{l-1}{k-1}$ gives the number of permutations in \eqref{eq:expansioncorrelationhierarchy} and $V_o(1):=B_d(R)\frac{N}{V}$ with $B_d(R)$ being the volume of a ball of radius $R$ in $d$ dimensions.

Inserting the expansion \eqref{eq:expansioncorrelationhierarchy} into the definition of the coefficients $D_l$, Eq.~\eqref{eq:definitionDcoefficients_supp}, we obtain analogously
\begin{align}
    D_l = \mu_{l-1} - \sum_{k=1}^{l-1} D_k V_o(l-k)\binom{l-1}{k-1},
    \label{eq:samplingDcoefficients_supp}
    \tag{$8$}
\end{align}
where
\begin{align}
    \mu_{l-1}&= N^{l-1}\int P_l \theta_{12} \dots \theta_{1l} \diff 1 \dots \diff l
    \notag
    \\
    &= N^{l-1}\int P_N \theta_{12} \dots \theta_{1l} \diff 1 \dots \diff N
\end{align}
is $N^{l-1}$ times the probability that particles $(2,\dots,l)$ are neighbors of particle one.
However this can be rewritten as
\begin{align}
    \mu_{l-1}&= N^{l-1} \sum_s p_n(s) \frac{1}{\binom{N-1}{s}} \binom{N-1-(l-1)}{s-(l-1)}
    \notag
    \\
    &=N^{l-1} \sum_s p_n(s) \frac{s!(N-1-s)!}{(N-1)!}
    \notag
    \\
    &\phantom{=N^{l-1} \sum_s }\times\frac{(N-l)!}{(N-s-1)!(s-l+1)!}
    \notag
    \\
    &= \sum_s p_n(s) \frac{s!}{(s-l+1)!}
\end{align}
which coincides with the previous definition of $\mu_l$, Eq.~\eqref{eq:defmu}.
In the first line we have the probability that there are $s$ neighbors of particle one times the probability that these neighbors are exactly $2, 3, \dots, l, n_1, n_2, \dots, n_{s-(l-1)}$ times the number of possible choices for $n_1, n_2, \dots, n_{s-(l-1)}$.

Eqs.~\eqref{eq:samplingCcoefficients_supp} and \eqref{eq:samplingDcoefficients_supp} are a system of linear equations relating the number of neighbor distribution parameters $C_l$ and $D_l$ to the measured quantities $V_o(l)$ and $\mu_l$.
We give the coefficients explicitly up to $l=5$ as
\begin{align}
    C_1=& B_d(R)\frac{N}{V},
    \notag
    \\
    C_2=& V_o(2)-C_1^2,
    \notag
    \\
    C_3=& V_o(3)-3C_1 V_o(2)+2C_1^3,
    \notag
    \\
    C_4=& V_o(4)-4C_1 V_o(3)- 3 V_o(2)^2 
    \notag
    \\
    &+12 C_1^2 V_o(2) -6 C_1^4,
    \notag
    \\
    C_5=& V_o(5)-5C_1 V_o(4)-16V_o(2)V_o(3) + 48 C_1 V_o(2)^2
    \notag
    \\
    &+20 C_1^2 V_o(3) -72 C_1^3V_o(2) +24 C_1^5
\end{align}
and
\begin{align}
    D_1=&1,
    \notag
    \\
    D_2=& \mu_1-C_1,
    \notag
    \\
    D_3=& \mu_2-V_o(2)-2C_1 \mu_1 + 2 C_1^2,
    \notag
    \\
    D_4=&\mu_3 - V_o(3) + 6 C_1 V_o(2) -3 \mu_1 V_o(2) + 6 \mu_1 C_1^2
    \notag
    \\
    &-3\mu_2 C_1 - 6C_1^3,
    \notag
    \\
    D_5=& \mu_4 -V_o(4)-4\mu_1 V_o(3) + 8 C_1 V_o(3) -12 \mu_2 V_o(2) 
    \notag
    \\
    &+12 V_o(2)^2 + 24 C_1\mu_1 V_o(2) -48 C_1^2V_o(2) 
    \notag
    \\
    &-4C_1\mu_3  -24 C_1^1\mu_1 +12 C_1^2 \mu_2 +24 C_1^4.
\end{align}
Note that $C_1$ and $D_1$ are not related to measured quantities but $D_1$ is trivially equal to one and $C_1$ is just the expected number of particles within a ball of radius $R$, hence it is just a system parameter.

\section{Overlap area of two and three disks\label{sec:overlap}}

\subsection{Two Disks}

\begin{figure}
    \centering
    \includegraphics[width=0.9\columnwidth]{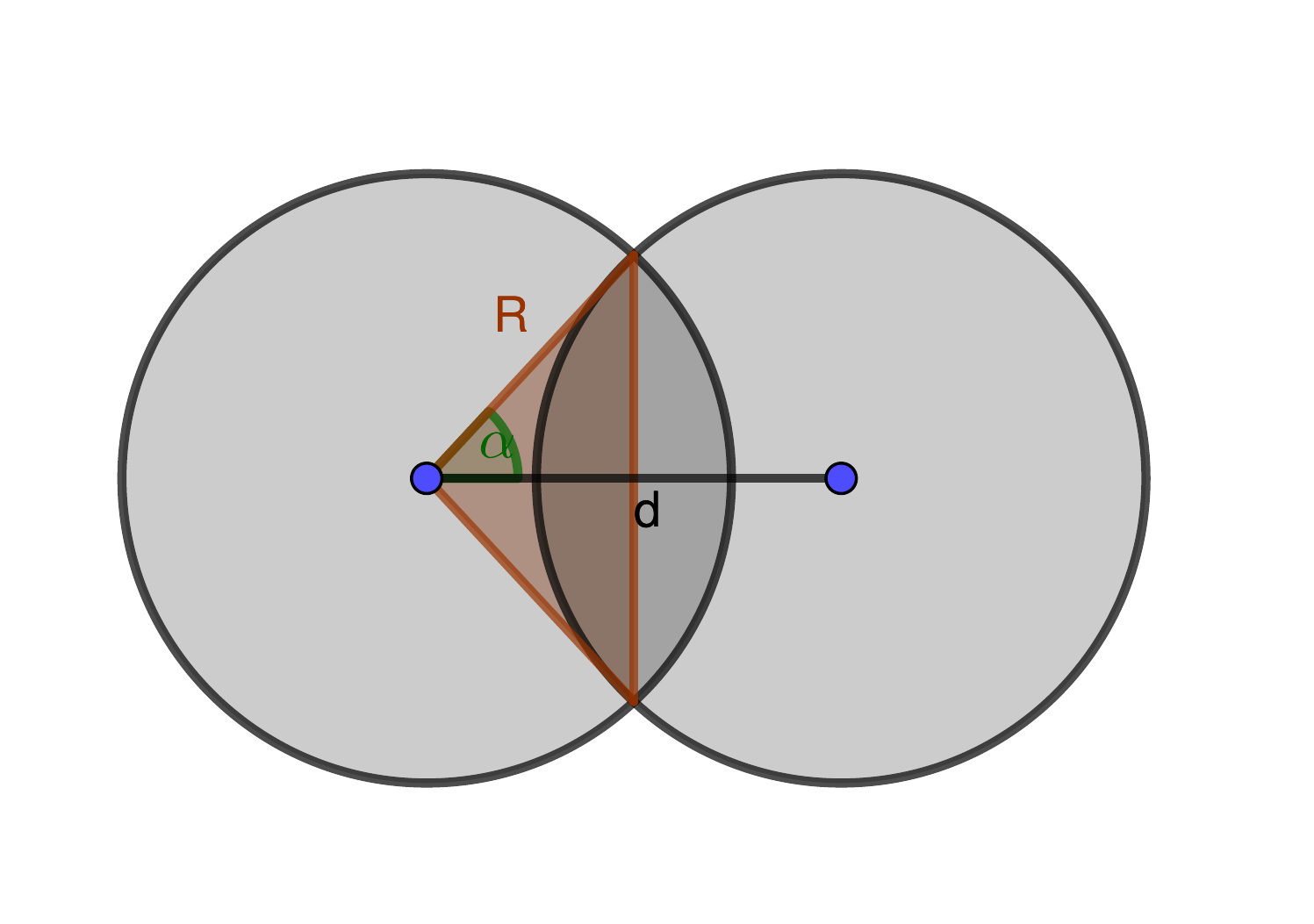}
    \caption{Overlap area of two disks. It can be calculated by taking twice the difference of areas of the circle segment with angle $2\alpha$ and of the triangle.}
    \label{fig:twodisksoverlap}
\end{figure}{}

For two disks of Radius $R$ and with distance $d$ between the centers of the disks the overlap area can be calculated as twice the difference of the circle segment and the triangle, see Fig.\ref{fig:twodisksoverlap} for a scetch.
The area of the circle segment is
\begin{align}
    A_{circle segment}= \alpha R^2
    \label{eq:circlesegment}
\end{align}
and the area of the triangle is given by 
\begin{align}
    A_{triangle}= \frac{d}{2} R\sin \alpha,
\end{align}
where the angle $\alpha$ is given by
\begin{align}
    \frac{d}{2}=R \cos \alpha.
\end{align}
Combining these expressions we find for the overlap area
\begin{align}
    A_{overlap}= 2R^2 \arccos{\bigg( \frac{d}{2R} \bigg) } - \frac{d}{2}\sqrt{4R^2-d^2}.
    \label{eq:twodiskoverlap}
\end{align}{}

\subsection{Three Disks}

We calculate the overlap area of three disks of radius $R$ with their centers pairwise separated by distances $a, b$ and $c$.
Clearly the overlap area only depends on those three distances.
Without loss of generality we assume for simplicity $a\ge b \ge c$.

We have to consider three different cases.

\subsubsection{Case A- no overlap}

\begin{figure}
    \centering
    \includegraphics[width=0.9\columnwidth]{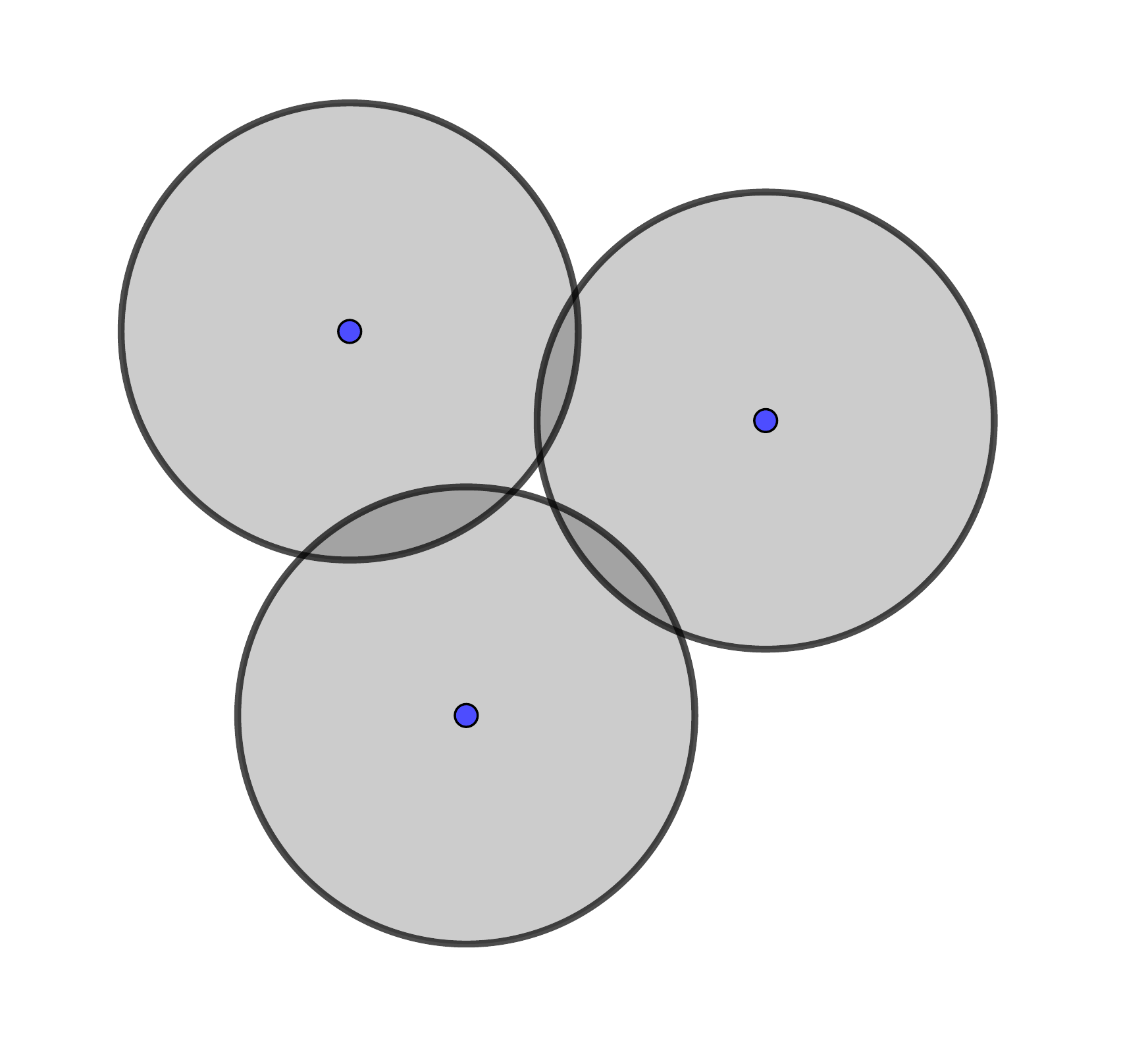}
    \caption{Case A - no overlap of three disks although there is a pairwise overlap for each pair.}
    \label{fig:case_a}
\end{figure}{}

Clearly, there is no overlap whenever $a \ge 2R$ since in that case the two circles separated by $a$ have zero overlap area.
However, even if the three disks have a pairwise overlap for all pairs, there might be still zero overlap of all three disks, see Fig.\ref{fig:case_a}.
Of cause in this case
\begin{align}
    A_{overlap}=0.
\end{align}{}

\subsubsection{Case B - two disk overlap}

\begin{figure}
    \centering
    \includegraphics[width=0.9\columnwidth]{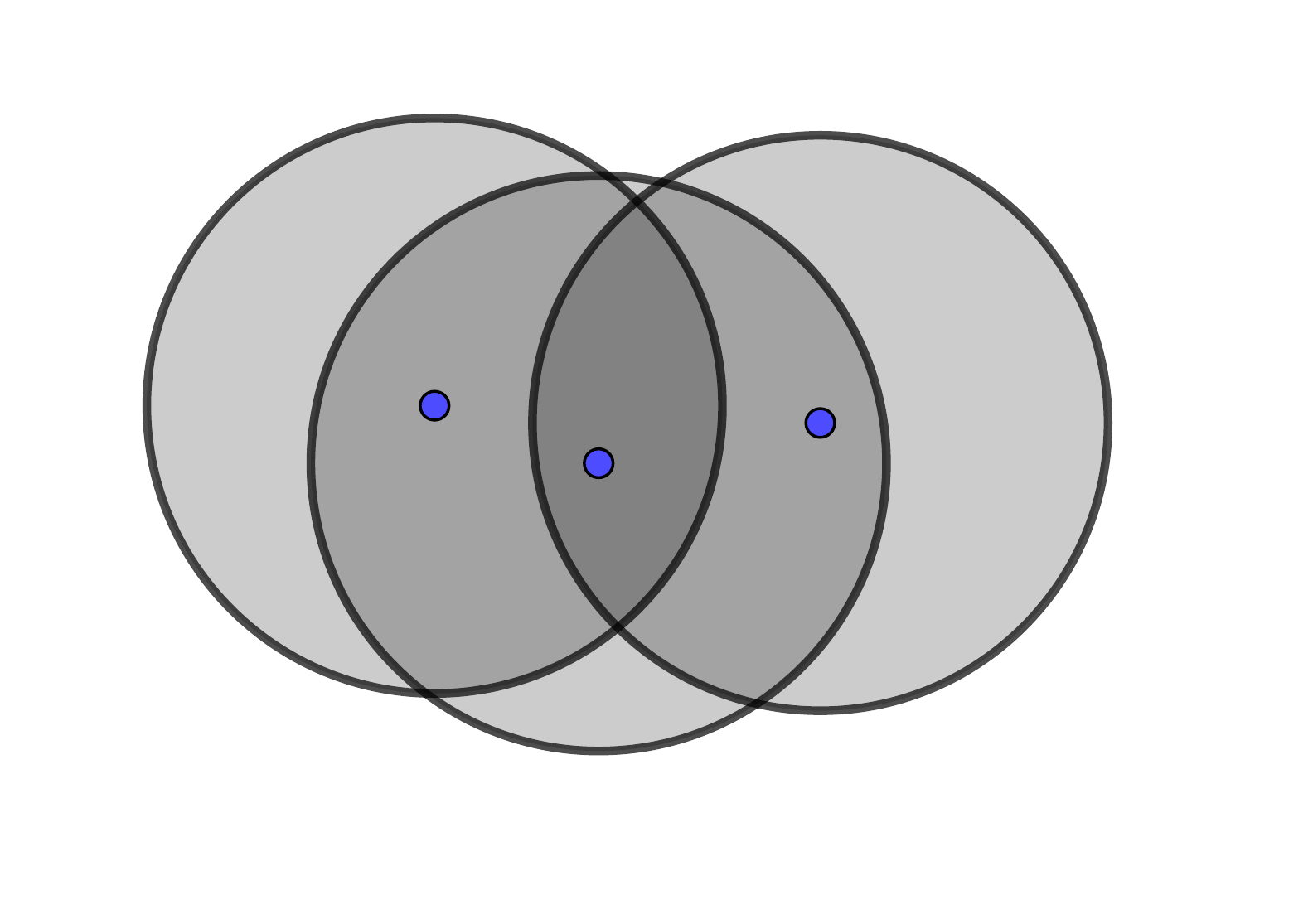}
    \caption{Case B - the overlap of three disks is identical to the overlap of two of the three disks.}
    \label{fig:case_b}
\end{figure}{}

It might happen that the three disk overlap is identical to the two disk overlap of the disks separated by $a$, see Fig.\ref{fig:case_b}.
In that case the overlap area is 
\begin{align}
    A_{overlap}= 2R^2 \arccos{\bigg( \frac{a}{2R} \bigg) } - \frac{a}{2}\sqrt{4R^2-a^2},
\end{align}{}
which is identical to Eq.~\eqref{eq:twodiskoverlap} with $d$ replaced by $a$.

\subsubsection{Case C - Reuleaux triangle like overlap}

\begin{figure}
    \centering
    \includegraphics[width=0.9\columnwidth]{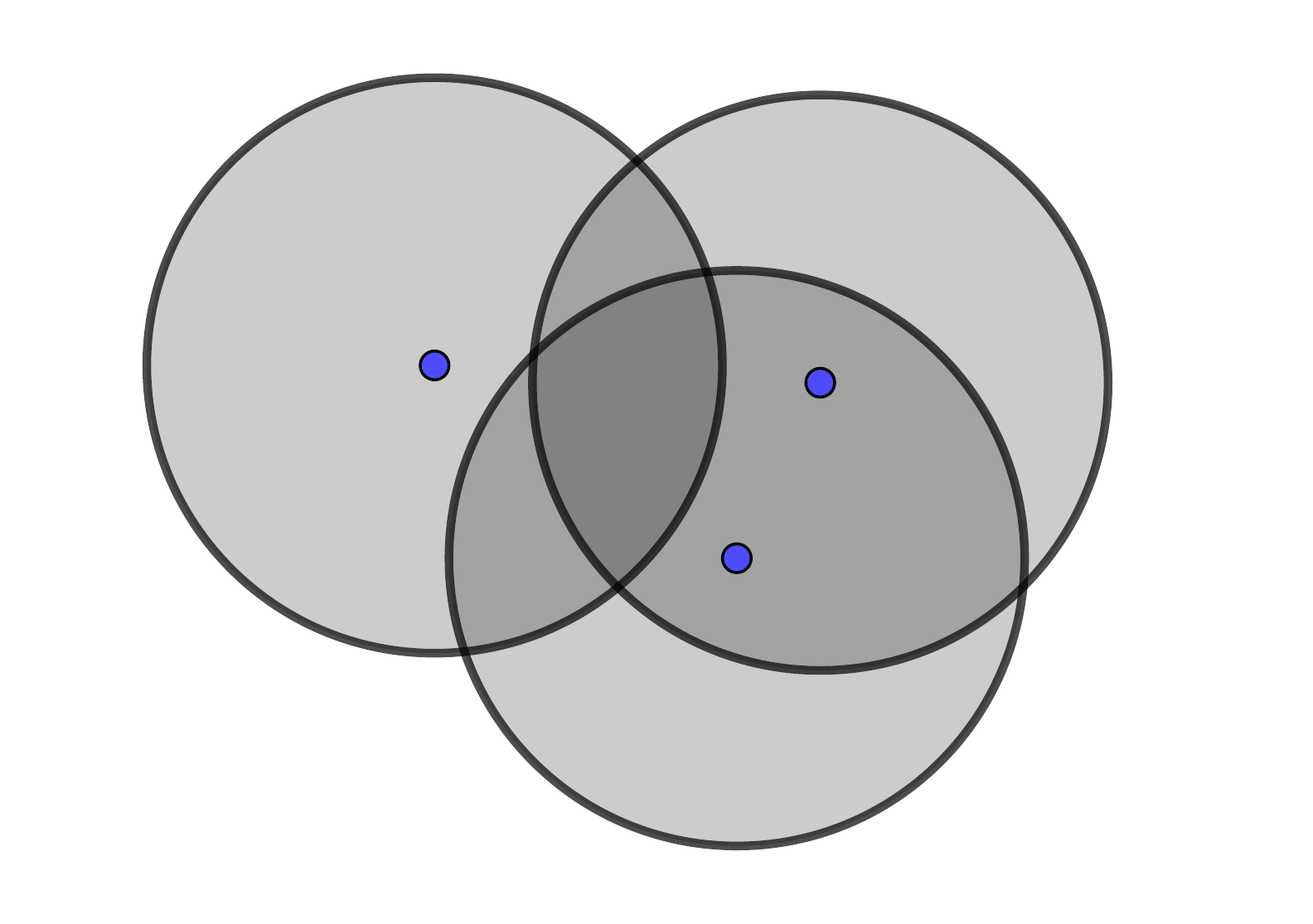}
    \caption{Case C - overlap of three disks - a generalization of the Reuleaux triangle.}
    \label{fig:case_c}
\end{figure}{}

In this most interesting case the overlap is nonempty and restricted by the boundaries of all three disks, see Fig.\ref{fig:case_c}.
In the special case of $a=b=c=R$ the overlap is a so called Reuleaux triangle which has important applications in arts, architecture and engineering.

Before calculating the overlap area in the third case it is important to find clear criteria, depending only on $a,b$ and $c$ that distinguish between the cases A, B and C.
Therefor it is useful to study the transitions of $A \leftrightarrow B$, $A \leftrightarrow C$ and $B \leftrightarrow C$ by moving one of the disks.

\begin{figure}
    \centering
    \includegraphics[width=0.9\columnwidth]{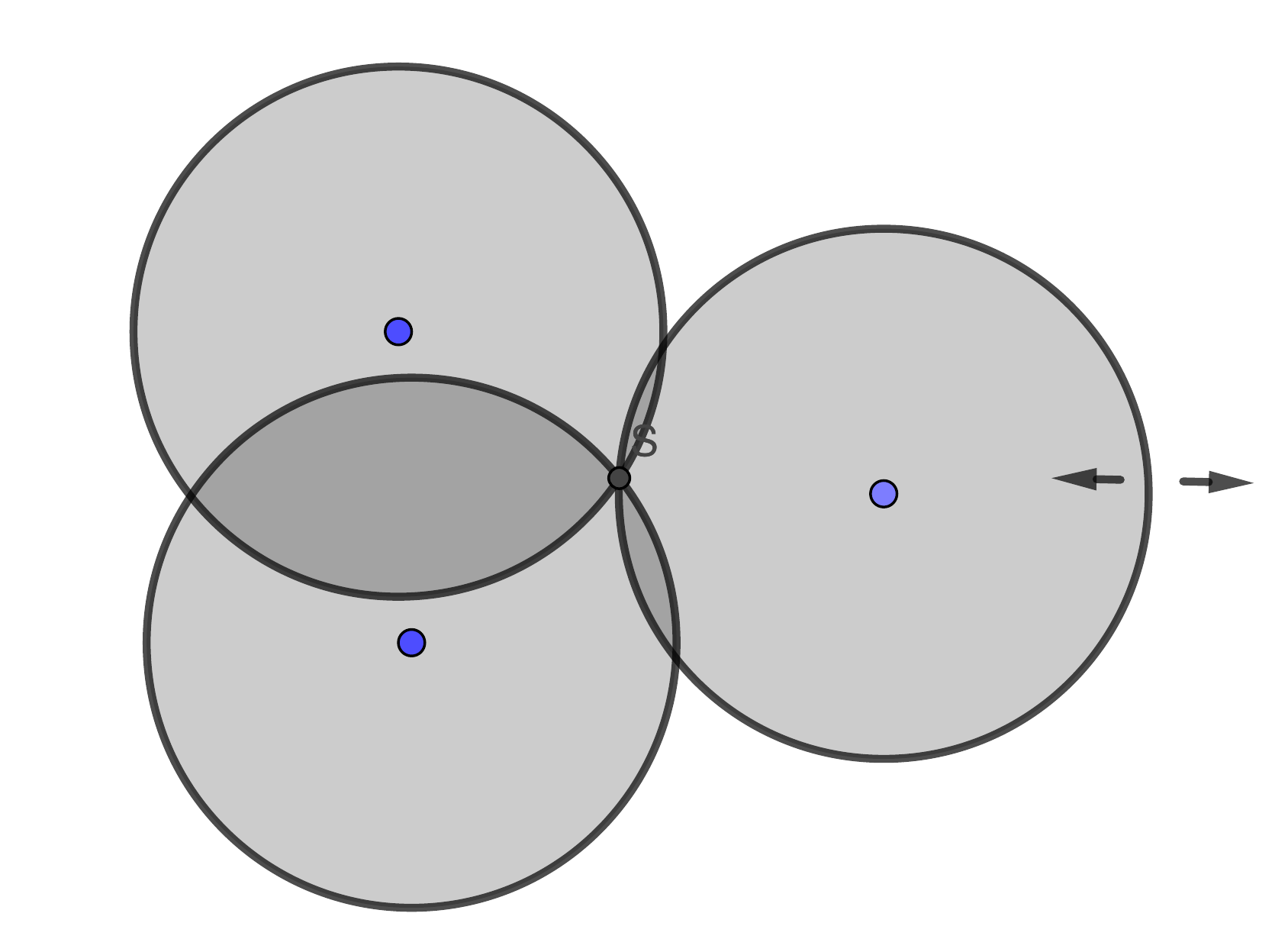}
    \caption{Transition between cases A and C. The overlap consists only out of the single point $S$. It is the center of the circumference of the centers of the disks.}
    \label{fig:transition_a_c}
\end{figure}{}

First, we consider the transition from C to A.
As soon as the overlap area has fallen to zero we have the situation sketched in Fig.\ref{fig:transition_a_c}.
The overlap consists only out of the single point $S$.
It has the same distance $R$ to the centers of the three disks.
Hence, $S$ is the center of the circumference of the triangle given by the centers of the disks and the circumference has radius $R$.
If one of the disks is now moved away from the overlap of the other two there is no overlap at all, we are in case A, and the radius of the circumference increases.
If on the other side one of the disks is moved towards the overlap of the other to, the three disks overlap becomes a real area, we are in case C, and the radius of the circumference decreases.

The transition $A \leftrightarrow B$ occurs when $a=2R$, however, me might just consider the case that $a<2R$, then this transition is not relevant.

\begin{figure}
    \centering
    \includegraphics[width=0.7\columnwidth]{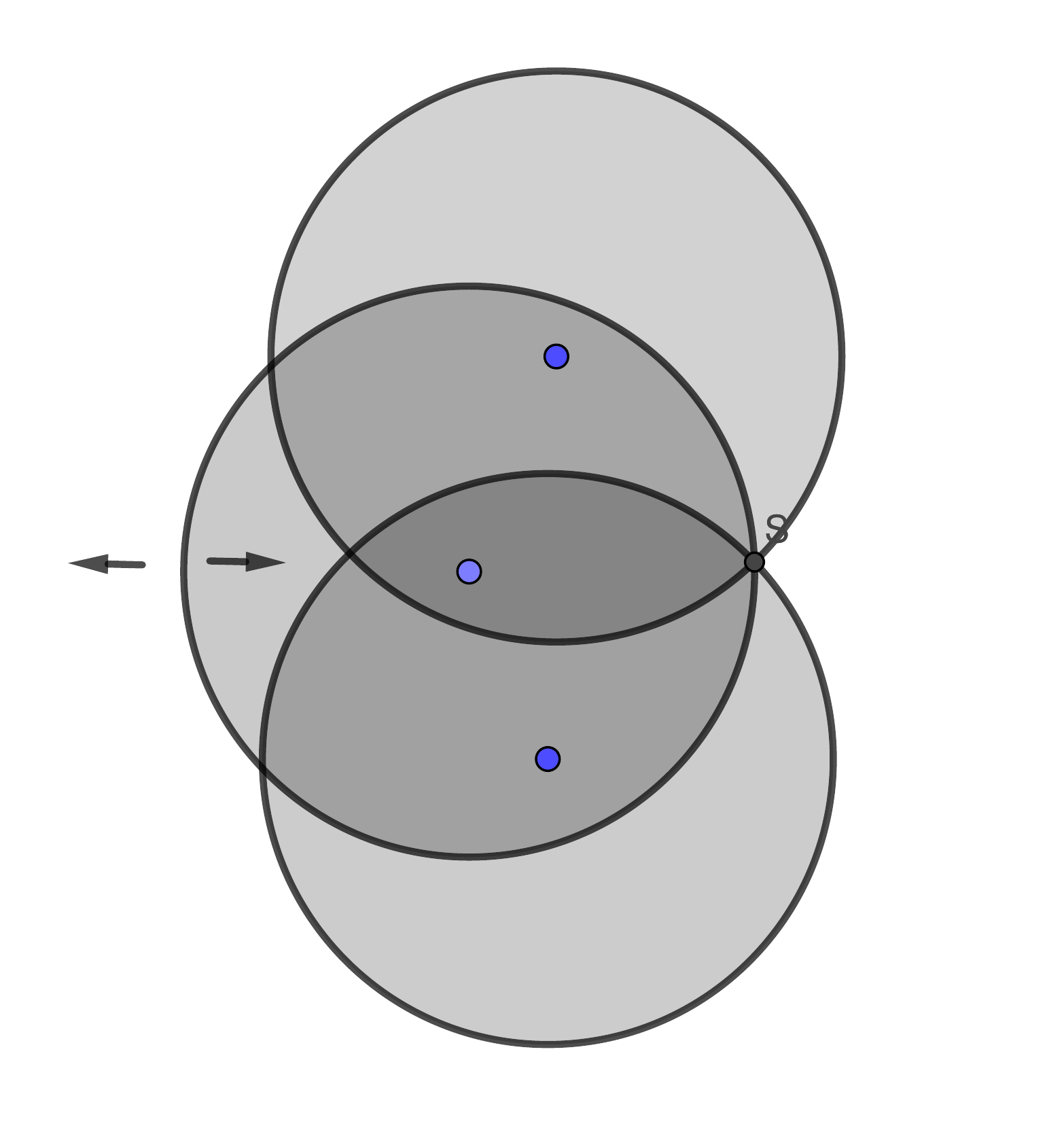}
    \caption{Transition between cases B and C. The overlap consists only out of the overlap of only two of the disks. The point $S$ is the center of the circumference of the centers of the disks.}
    \label{fig:transition_b_c}
\end{figure}{}

The Transition $B \leftrightarrow C$ is sketched in Fig.\ref{fig:transition_b_c}.
Again, the transition occurs when the three circles intersect in just one point $S$ which has the same distance $R$ to the centers of all three disks and thus $R$ is the circumference diameter.
If one of the disks is moved either the overlap becomes just the two circle overlap, we are in case B, and the circumference radius increases or the overlap becomes a Reuleaux-like triangle, we are in case C, and the circumference radius decreases.

Thus we found that case C occurs whenever for the circumference radius it holds
\begin{align}
    r_{circumference}=\frac{abc}{4\sqrt{s(s-a)(s-b)(s-c)}}<R
\end{align}{}
with $s=(a+b+c)/2$.

Whenever $r_{circumference}\ge R$ we still need to distinguish between cases A and B.
As we assumed $a<2R$ the midpoint $M_a$ of the two disk centers separated by $a$ is always in the overlap of those two disks.
If we draw a circle with diameter $a$ around $M_a$ and the center of the third disk is inside this circle it follows that the three-disk overlap is nonempty and thus we are in case B. 
Thales's theorem tells us that in this case the triangle given by $a, b, c$ is obtuse. 
Hence by Pythagoras's theorem we conclude that $a^2>b^2+c^2$ in this case.
If on the other hand the center of the third disk lies outside the circle we conclude that $M_a$ is not in the three-disk overlap and hence we can not be in case B. 
Thus we must be in case A.
Again by Thales's theorem the triangle given by $a,b,c$ must be acute in this case and therefore by Pythagoras's theorem $a^2<b^2+c^2$.

In summary, we can distinguish the cases A, B and C by
\begin{align}
    \text{Case A }&\leftrightarrow r_{circumference}\ge R \text{ and } a^2<b^2+c^2, 
    \notag
    \\
    \text{Case B }&\leftrightarrow r_{circumference}\ge R \text{ and } a^2>b^2+c^2, 
    \notag
    \\
    \text{Case C }&\leftrightarrow r_{circumference}< R.  
\end{align}{}

\begin{figure}
    \centering
    \includegraphics[width=0.7\columnwidth]{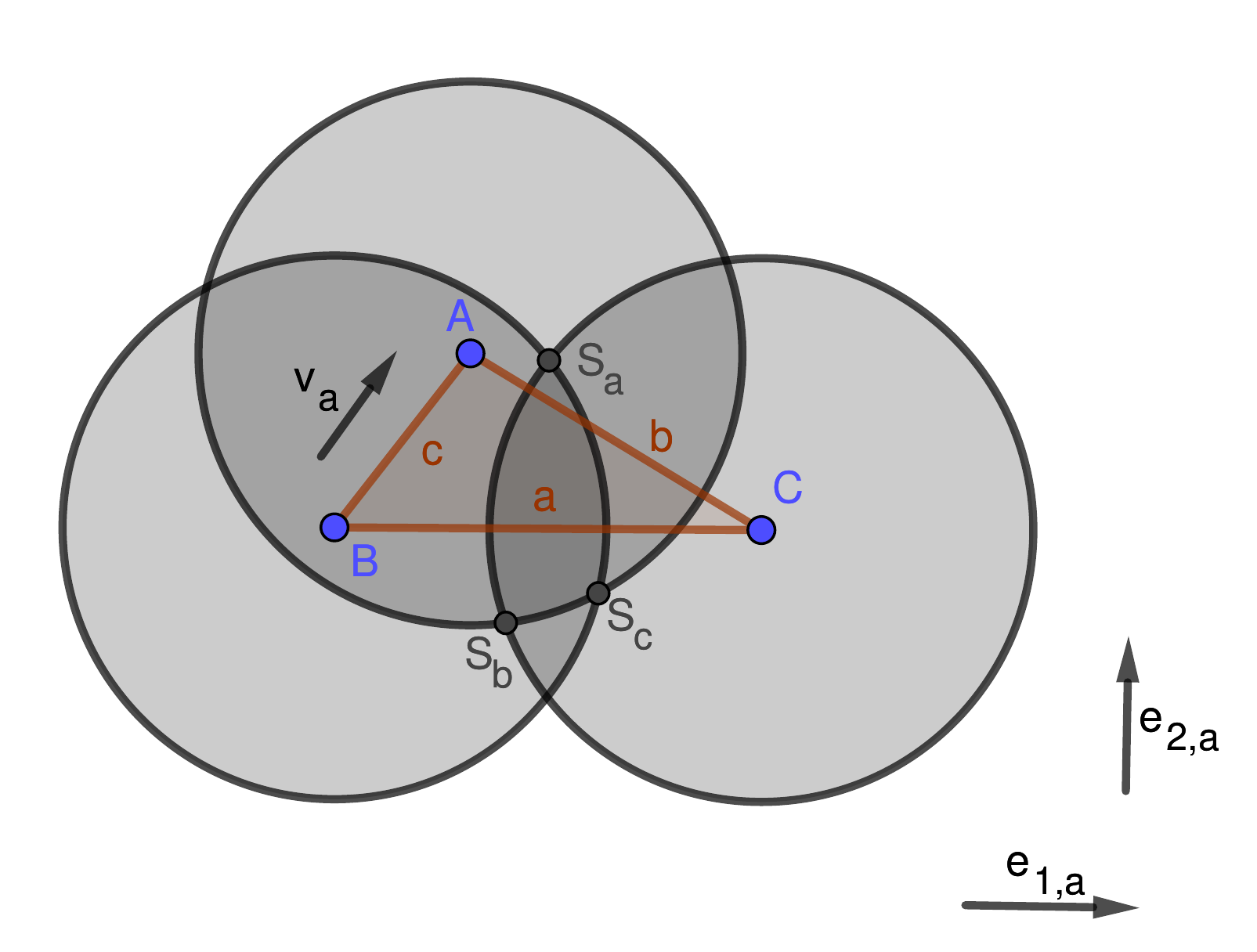}
    \caption{Sketch of the Reuleaux-like overlap.}
    \label{fig:reuleaux1}
\end{figure}{}

It remains to calculate the overlap area in case C.
We start to calculate the distances $l_1, l_2, l_3$ between the corners of the overlap area.
We use the carthesian coordinates
\begin{align}
    \binom{x_1}{y_1}=B, \, \, \binom{x_2}{y_2}=C, \, \, \binom{x_3}{y_3}=A
\end{align}{}
for the centers of the three disks. Hence we find for the distances between the disk centers
\begin{align}
    a=&\sqrt{(x_2-x_1)^2+(y_2-y_1)^2}
    \notag
    \\
    b=&\sqrt{(x_3-x_2)^2+(y_3-y_2)^2}
    \notag
    \\
    c=&\sqrt{(x_1-x_3)^2+(y_1-y_3)^2}.
\end{align}{}
We define the following unit vectors.
$\vec{e}_{1,a}$ points from $\binom{x_1}{y_1}$ to $\binom{x_2}{y_2}$, $\vec{e}_{1,b}$ points from $\binom{x_2}{y_2}$ to $\binom{x_3}{y_3}$ and $\vec{e}_{1,c}$ points from $\binom{x_3}{y_3}$ to $\binom{x_1}{y_1}$.
Furthermore we define the vectors $\vec{v}_i$ via
\begin{align}
    \vec{v}_a&= \binom{x_3-x_1}{y_3-y_1}
    \notag
    \\
    \vec{v}_b&= \binom{x_1-x_2}{y_1-y_2}
    \notag
    \\
    \vec{v}_c&= \binom{x_2-x_3}{y_2-y_3}.
\end{align}{}
The unit vectors $\vec{e}_{2,i}$ are defined by being perpendicular to the corresponding unitvector $\vec{e}_{1,i}$ and the property $\vec{v}_i \cdot \vec{e}_{2,i}>0$, where $i=a,b,c$, see Fig.~\ref{fig:reuleaux1} for a sketch of the unit vectors.
They can be explicitly calculated by
\begin{align}
    \vec{e}_{2,i}= \frac{\vec{v}_i-(\vec{v}_i\cdot \vec{e}_{1,i}) \vec{e}_{1,i} }{|\vec{v}_i-(\vec{v}_i\cdot \vec{e}_{1,i}) \vec{e}_{1,i}|}.
    \label{eq:definitionunitvectortwo}
\end{align}{}
It is a simple geometric consideration to calculate $S_a$ as 
\begin{align}
    S_a= M_a + \sqrt{R^2-a^2/4}\vec{e}_{2,a},
\end{align}{}
where $M_a$ is the midpoint of the side $a$, that is $M_a=\frac{1}{2}(B+C)= \frac{1}{2}\binom{x_1+x_2}{y_1+y_2}$.
Thus we find the intersection point between the circles $S_a$ and analogously $S_b$ as
\begin{align}
    S_a&= \frac{1}{2} \binom{x_1+x_2}{y_1+y_2} + \sqrt{R^2- \frac{a^2}{4}}\vec{e}_{2,a},
    \notag
    \\
    S_b&= \frac{1}{2} \binom{x_2+x_3}{y_2+y_3} + \sqrt{R^2- \frac{b^2}{4}}\vec{e}_{2,b}.
    \label{eq:intersectionpoints}
\end{align}{}

\begin{figure}
    \centering
    \includegraphics[width=0.7\columnwidth]{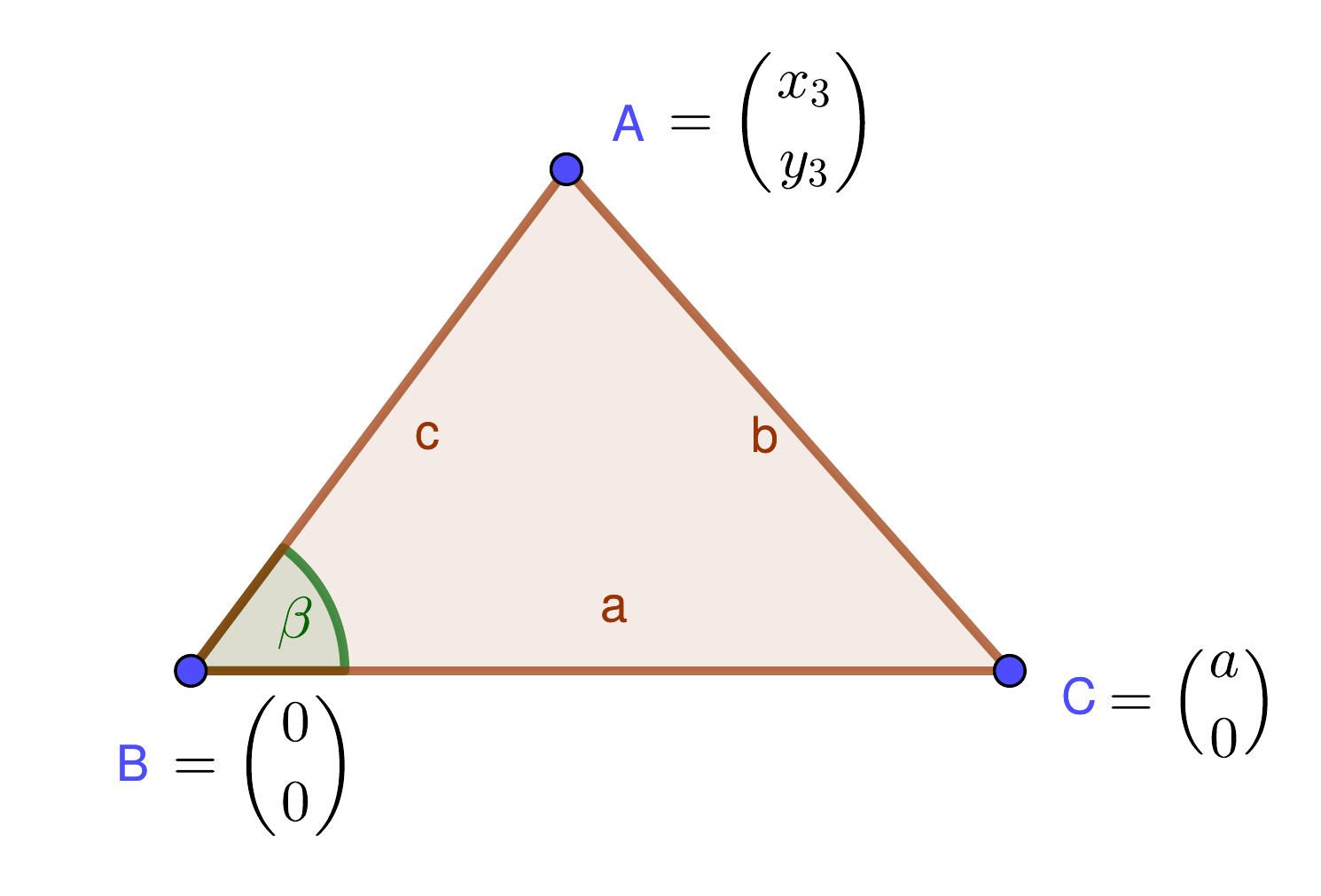}
    \caption{We choose a coordinate system such that $B=\binom{0}{0}$, $C=\binom{a}{0}$ and $A=\binom{x_3}{y_3}$, with $x_3,y_3>0$.}
    \label{fig:reuleaux2}
\end{figure}{}

We are free to choose a special coordinate system such that $B=\binom{0}{0}$, $C=\binom{a}{0}$ and $A=\binom{x_3}{y_3}$ with $x_3,y_3>0$, see Fig.~\ref{fig:reuleaux2} for a sketch. Once $B$ and $C$ are fixed, the condition $x_3>0$ is one of two choices, whereas the condition $y_3>0$ follows from $a\ge b \ge c$.
We us the law of cosines to calculate the angle $\beta = \sphericalangle CBA$ and find 
\begin{align}
    \cos \beta &= \frac{a^2+c^2-b^2}{2ac} ,
    \notag
    \\
    \sin \beta &= \frac{\sqrt{2a^2b^2 +2a^2c^2+2b^2c^2-a^4-b^4-c^4}}{2ac}
\end{align}{}
and hence we found all coordinates of the centers of the disks
\begin{align}
x_1&=0
\notag
\\
y_1&=0
\notag
\\
x_2&=a
\notag
\\
y_2&=0
\notag
\\
    x_3&= c \cos \beta = \frac{a^2+c^2-b^2}{2a} ,
    \notag
    \\
    y_3&= c\sin \beta = \frac{\sqrt{2a^2b^2 +2a^2c^2+2b^2c^2-a^4-b^4-c^4}}{2a}.
    \label{eq:coordinates}
\end{align}{}
From our choice of the coordinate system we find
\begin{align}
    \vec{e}_{1,a}=& \binom{1}{0},
    \notag
    \\
    \vec{e}_{2,a}=& \binom{0}{1}.
    \label{eq:unitvectorsa}
\end{align}{}
As we explicitly have the coordinates of $C, A$ we can calculate $\vec{e}_{1,b}$ from its definition and $\vec{e}_{2, b}$ according to Eq.~\eqref{eq:definitionunitvectortwo} which results in
\begin{align}
    \vec{e}_{1,b} &= \frac{1}{2ab} \binom{c^2-a^2-b^2}{\sqrt{2a^2b^2+2a^2c^2+2b^2c^2-a^4-b^4-c^4}},
    \notag
    \\
    \vec{e}_{2, b} &= \frac{1}{2ab}\binom{-\sqrt{2a^2b^2+2a^2c^2+2b^2 c^2-a^4-b^4-c^4} }{c^2-a^2- b^2}.
    \label{eq:unitvectorsb}
\end{align}{}
Plugging Eqs.~\eqref{eq:coordinates}, \eqref{eq:unitvectorsa} and \eqref{eq:unitvectorsb} into Eq.~\eqref{eq:intersectionpoints} we find for two of the intersection points
\begin{align}
    &S_{a,x}= \frac{a}{2},
    \notag
    \\
    &S_{a,y}= \sqrt{R^2-\frac{a^2}{4}},
    \notag
    \\
    &S_{b,x}=\frac{a}{2} + \frac{a^2+c^2-b^2}{4a} 
    \notag
    \\
    &- \sqrt{R^2-\frac{b^2}{4}}\frac{\sqrt{2a^2b^2+2a^2c^2+2b^2c^2-a^4-b^4-c^4}}{2ab},
    \notag
    \\
    &S_{b,y}=\sqrt{R^2-\frac{b^4}{4}}\frac{c^2-b^2-a^2}{2ab}
    \notag
    \\
    &+\frac{\sqrt{2a^2b^2+2a^2c^2+2b^2c^2-a^4-b^4-c^4}}{4a}.
\end{align}{}
Eventually we can calculate the distance between the points $S_a$ and $S_b$ to find
\begin{align}
    l_1:&= \sqrt{(S_{a,x}-S_{b,x})^2+ (S_{a,y}-S_{b,y})^2}
    \notag
    \\
    &= \bigg[
- \frac{\sqrt{2 a^2  b^2 + 2  a^2 c^2 + 2  b^2  c^2 - a^4 -b^4 -c ^4}}{2} 
\notag
\\
&\times \bigg(\frac{\sqrt{R^2-\frac{a^2}{4}}}{a}+\frac{\sqrt{R^2-\frac{b^2}{4}}}{b}\bigg) + \frac{c^2 - b^2 -a^2}{4}
\notag
\\
&- 2  \sqrt{R^2 -\frac{a^2 }{4}}  \sqrt{R^2 -\frac{b^2 }{4}} \frac{c^2 - b^2 -a^2}{2  a  b} +2R^2\bigg]^{1/2}.
\end{align}{}
The other two lengths $l_2$ and $l_3$ are obtained by permutations of $a, b$ and $c$.

It remains to calculate the overlap are.
It consists out of a triangle with side lengths $l_1, l_2, l_3$ and three circular caps.
The triangle area is calculated by Heron's formula, whereas the circular caps can be calculated analogously to the two-particle overlap.
Putting everything together we obtain the overlap area 
\begin{align}
	&\mathcal{A}_{\text{overlap}} = \arcsin \left(\frac{l_1}{2 R}\right) \cdot R^2 - \frac{l_1}{2}  \sqrt{ R^2 -\frac{l_1^2}{4}} 
	\notag
	\\
	&+ \arcsin \left(\frac{l_2}{2 R}\right)  R^2 - \frac{l_2}{2} \sqrt{ R^2 -\frac{l_2^2}{4}} + \arcsin \left(\frac{l_3}{2 R}\right) R^2
	\notag
	\\
	&- \frac{l_3}{2} \sqrt{ R^2 -\frac{l_3^2}{4}} + \sqrt{l (l-l_1)  (l-l_2) (l-l_3)}.
	\label{eq:difficultcase}
\end{align}

\section{Models\label{sec:models} and numerical results}

\subsection{Vicsek model}

In the letter we present numerical results of the standard two dimensional Vicsek model.
In this model $N$ particles move at constant speed $v$ in individual directions given by angles $\theta_i$, that is
\begin{align}
    x_j(t+1)= x_j(t) + v \cos \theta_j(t),
    \notag
    \\
    y_j(t+1)= y_j(t) + v \sin \theta_j(t).
\end{align}{}
where $x_j(t), y_j(t)$ define the position of particle $j$ at time $t$.
After each unit of time the particles instantaneously change their directions according to the following interaction rule
\begin{align}
    \theta_j(t+1)= \arg \bigg\{\sum_{k \in \Omega_j(t+1)} \exp[i \theta_k(t)] \bigg\} + \eta \xi_i(t),
\end{align}{}
where $\Omega_j$ is the set of indexes of particles that are within distance $R$ of particle $j$ such that $\sqrt{(x_k-x_j)^2+(y_k-y_j)^2}\le R$ for all $k\in \Omega_j$.
We call the particles given by $\Omega_j$ the neighbors of particle $j$.
Note that it is always $j\in \Omega_j$.
The noise strength is given by $\eta \in [0,1]$ and $\xi_j(t)$ are independent random variables uniformly distributed on the interval $[-\pi, \pi]$.
It should be noted that the time interval of free motion between the collisions is sometimes considered as a system parameter $\tau$.
However we only use $\tau=1$ as the only relevant parameter is given by the product $\tau \cdot v$ anyways.
Periodic boundary conditions are used for the coordinates $x_i, y_i \in [-L/2,L/2)$ and $\phi_i \in [0, 2\pi)$. 

Zero noise $\eta=0$ leads to a completely deterministic motion, where particles align consequently and eventually all move in the same direction for almost all initial conditions.
On the other hand, all particles move in random directions for $\eta=1$.
To distinguish these cases it is useful to introduce the two-dimensional polar order parameter $\mathbf{p}$ with $|\mathbf{p}|\in[0,1]$ defined by
\begin{align}
    p_x = \frac{1}{N}\sum_{j=1}^N \cos \theta_j,
    \\
    p_y = \frac{1}{N}\sum_{j=1}^N \sin \theta_j.
\end{align}
Thus $|\mathbf{p}|=1$ corresponds to the collective motion of all particles in the same direction and $|\mathbf{p}|=0$ means that all particles move independently in random directions.

\begin{figure}
    \centering
    \includegraphics[width=0.7\columnwidth]{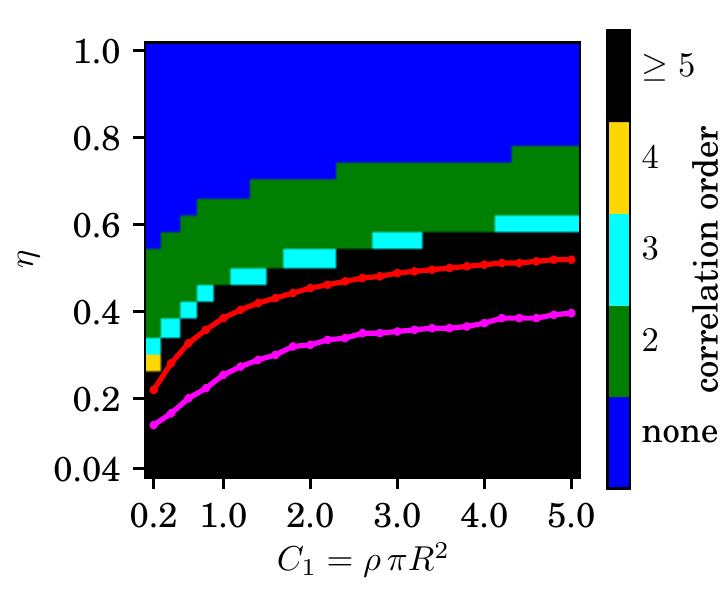}
    \caption{Minimal required correlation order as in Fig.~2 of the Letter but with a threshold of $10^{-2}$ for the Kullback-Leibler divergence.}
    \label{fig:corr_order_limit2}
\end{figure}{}

\begin{figure}
    \centering
    \includegraphics[width=0.7\columnwidth]{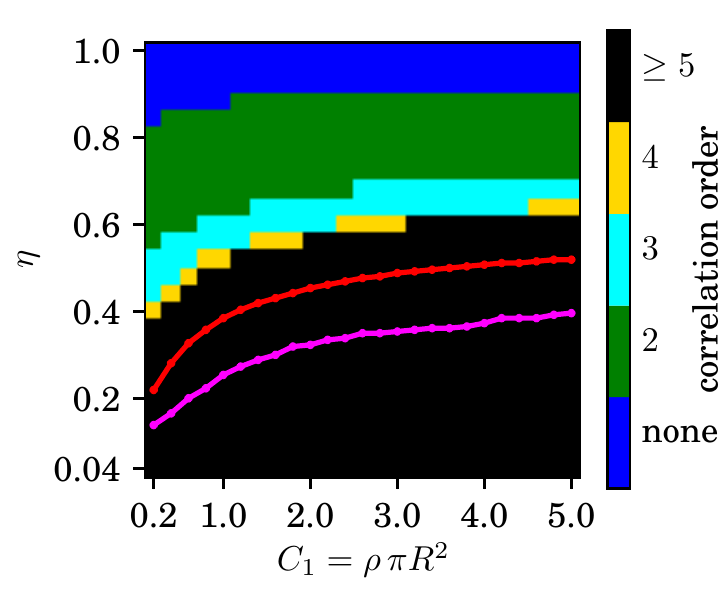}
    \caption{Minimal required correlation order as in Fig.~2 of the Letter but with a threshold of $10^{-4}$ for the Kullback-Leibler divergence.}
    \label{fig:corr_order_limit3}
\end{figure}{}

\begin{figure}
    \centering
    \includegraphics[width=0.7\columnwidth]{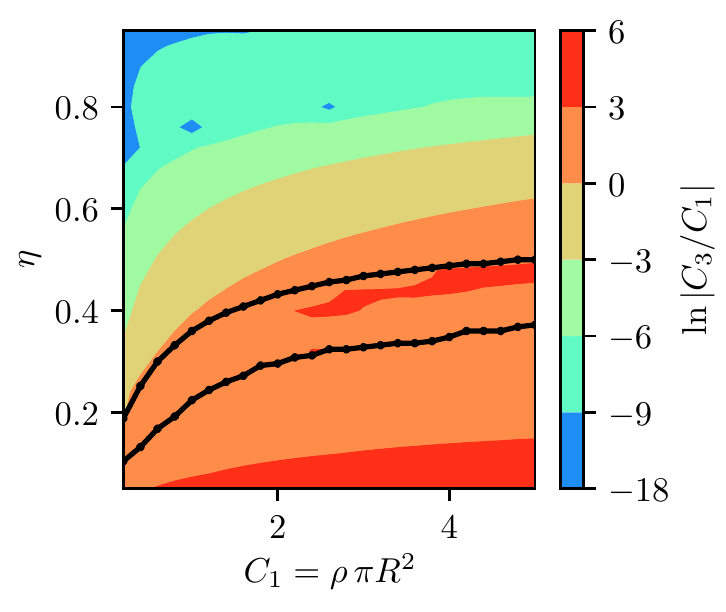}
    \caption{Ratio of three particle correlation parameter $C_3$ and mean number of neighbors $C_1$ for parameters as in Fig. 3 of the Letter. Black lines show the transition lines. For large noise, when $C_3$ is very small its measured value is sometimes negative (not shown).}
    \label{fig:cthree}
\end{figure}{}

\begin{figure}
    \centering
    \includegraphics[width=0.7\columnwidth]{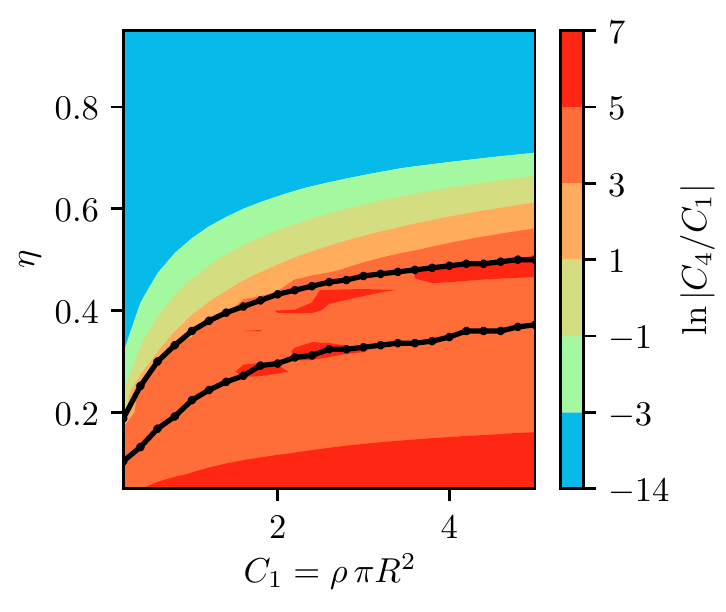}
    \caption{Ratio of four particle correlation parameter $C_4$ and mean number of neighbors $C_1$ for parameters as in Fig. 3 of the Letter. Black lines show the transition lines. For large noise, when $C_4$ is very small its measured value is sometimes negative (not shown).}
    \label{fig:cfour}
\end{figure}{}

In Figs.~\ref{fig:corr_order_limit2} and \ref{fig:corr_order_limit3} we show the \textit{correlation map} equivalent to Fig.~2 of the Letter, but obtained using a different threshold for the Kullback-Leibler divergence of $10^{-2}$ and $10^{-4}$ in Figs.~\ref{fig:corr_order_limit2} and \ref{fig:corr_order_limit3}, respectively. 
For these other threshold values of the Kullback-Leibler divergence the correlation orders are slightly shifted.
However, the over all picture remains unchanged.

In Figs.~\ref{fig:cthree} and \ref{fig:cfour} we show the three- and four-particle correlation parameters obtained from the same simulations as we used for Fig.~3 of the Letter. They look qualitatively similar to the two-particle correlation parameter $C_2$.

\begin{figure}
    \centering
    \includegraphics[width=0.7\columnwidth]{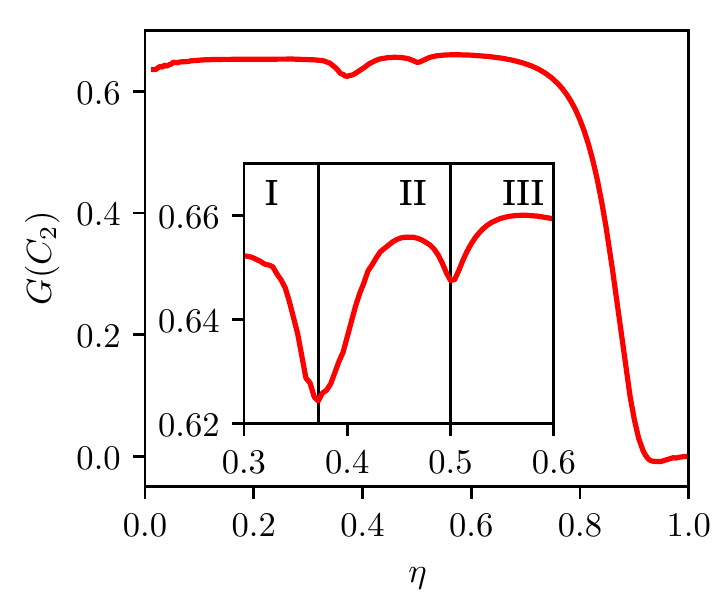}
    \caption{Binder cumulant of the two-particle correlation parameter $C_2$ clearly showing two transitions. System parameters: $N=22500$, $v\tau/R=1$, $C_1=5$, thermalization for $10^5$ and measurement for $10^6$ time steps for 24 realizations.}
    \label{fig:binder2}
\end{figure}

\begin{figure}
    \centering
    \includegraphics[width=0.7\columnwidth]{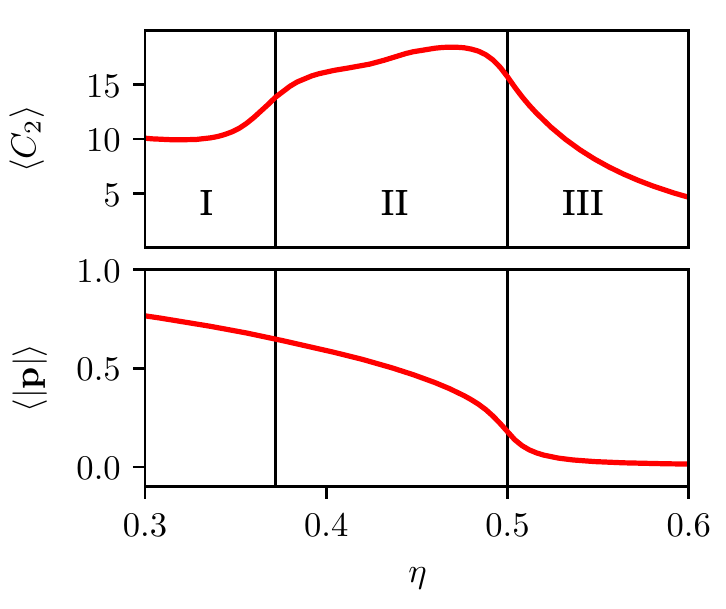}
    \caption{\textit{Top:} Two-particle correlation parameter $C_2$. \textit{Bottom:} Polar order parameter. Data from same simulations as Fig.~\ref{fig:binder2}. Vertical lines show transitions obtained as minima of the Binder cumulant in Fig.~\ref{fig:binder2}.
    \label{fig:ctwo_polar}}
\end{figure}

\begin{figure}
    \centering
    \includegraphics[width=0.7\columnwidth]{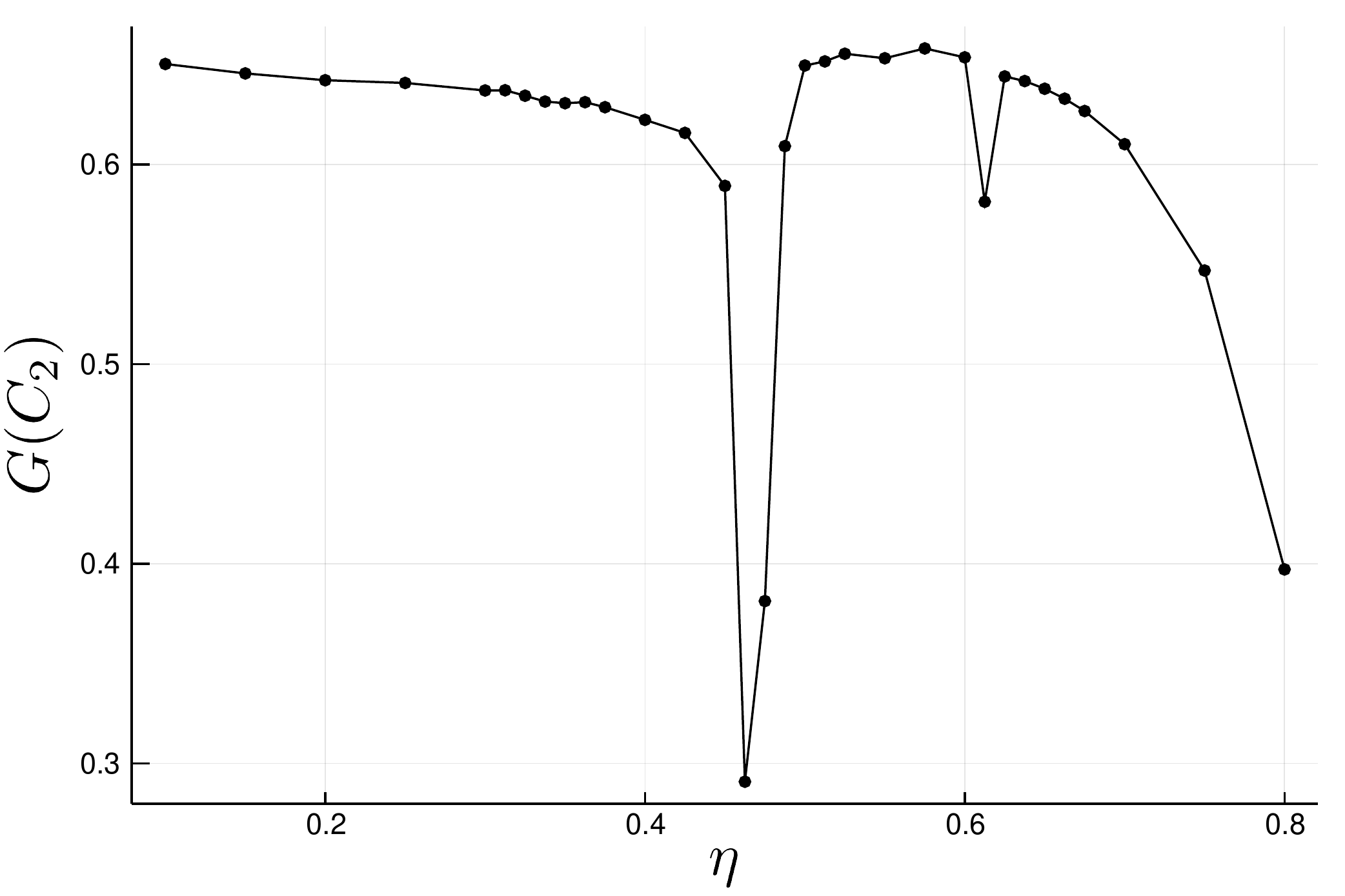}
    \caption{Binder cumulant of the two-particle correlation parameter $C_2$ clearly showing two transitions. System parameters: $N=22500$, $v\tau/R= 5$, $C_1=5$, thermalization for $10^5$ and measurement for $10^6$ time steps for 25 realizations.}
    \label{fig:binder3}
\end{figure}

In Figs.\ref{fig:binder2} and \ref{fig:ctwo_polar} we display the two-particle correlation parameter $C_2$, its Binder cumulant and the polar order parameter analogously to Fig.~4 of the Letter for another parameter set. Here we see even clearer minima in the Binder Cumulant. In Fig.~\ref{fig:binder3} we display the Binder cumulant for another parameter set (large speed). There, we also find clearly two transitions.

\begin{figure}
    \centering
    \includegraphics[width=0.7\columnwidth]{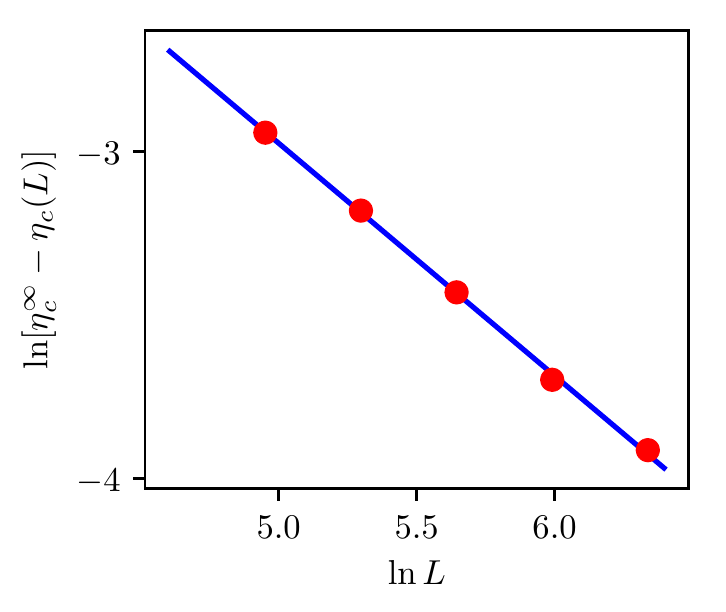}
    \caption{Finite size scaling for the critical noise strength of the polar order transition of the simulations presented in Fig.~4 of the Letter. The blue line represents a fit of the finite size scaling ansatz \eqref{eq:finitesizescalingansatz}. The red circles show the measured values for different system sizes. 
    \label{fig:finite_size_scaling}}
\end{figure}

We obtain the critical point $\eta_c(L)$ for the polar order transition as the right local minimum of the Binder cumulant in Fig.~4 (c) of the Letter for different finite system sizes $L$. To extrapolate to infinitely large systems we use the finite size scaling ansatz
\begin{align}
    \eta_c(L)= \eta_c^\infty -A\cdot L^\alpha,
    \label{eq:finitesizescalingansatz}
\end{align}
If the wave fronts occuring in the polar ordered phase would have the same shape for all system sizes and noise strengths, we would require that the exponent $\alpha$ equals to minus one. 
However, the shape of the wave fronts depends on the noise strength.
Therefore, we need to fit all three parameters and obtain the values $\eta_c^\infty=0.285$, $A=1.79$ and $\alpha=-0.711$.
The ansatz \eqref{eq:finitesizescalingansatz} is shown together with the measured values of $\eta_c(L)$ in Fig.~\ref{fig:finite_size_scaling}.
The results seem to agree very well, however, we should keep in mind that we fitted three parameters with five data points.
In order to give a quantitative uncertainty estimate of the fitted parameters, we need data for more and larger system sizes.
The results deliver therefore only a rough estimate of the critical noise strength for the infinite system size.
However, the value we obtain via finite size scaling is already quite close to the value that was obtained by using a different method (values given in the Letter).

\subsection{Continuous time Vicsek-like model}

Here, we consider a model similar to the Vicsek model but with continuous time. 
As in the standard Vicsek model particles move in two dimensions with constant speed and align their direction of motion with nearby particles.
The system is defined by the following Langevin equations
\begin{align}
	\dot{x}_{i} &= v \cos(\theta_{i}) 
	\notag
	\\
	\dot{y}_{i} &= v \sin(\theta_{i}) 
	\notag
	\\
	\dot{\theta}_{i} &= w(|\Omega(i)|) \sum_{j\in \Omega(i)} \sin(\theta_j-\theta_i) + \sigma \xi_{i}, \quad i=1, \dots, N,
	\label{eq:model}
\end{align}
Here, the neighborhood sets $\Omega_i$ are defined as in the standard Vicsek model above.
The $\xi_{i}(t)$ are independent Gaussian white noise terms satisfying 
\begin{align}
	\langle \xi_{i}(t) \xi_j(s) \rangle = \delta_{ij}\delta(t-s).
	\label{eq:whitenoise}
\end{align}
The interaction weight $w$ is a function of the number of neighbors including the particle itself.
In this paper we consider the case $w(n)=1/n$. 

\begin{figure}[H]
    \centering
    \includegraphics[width=0.9\columnwidth]{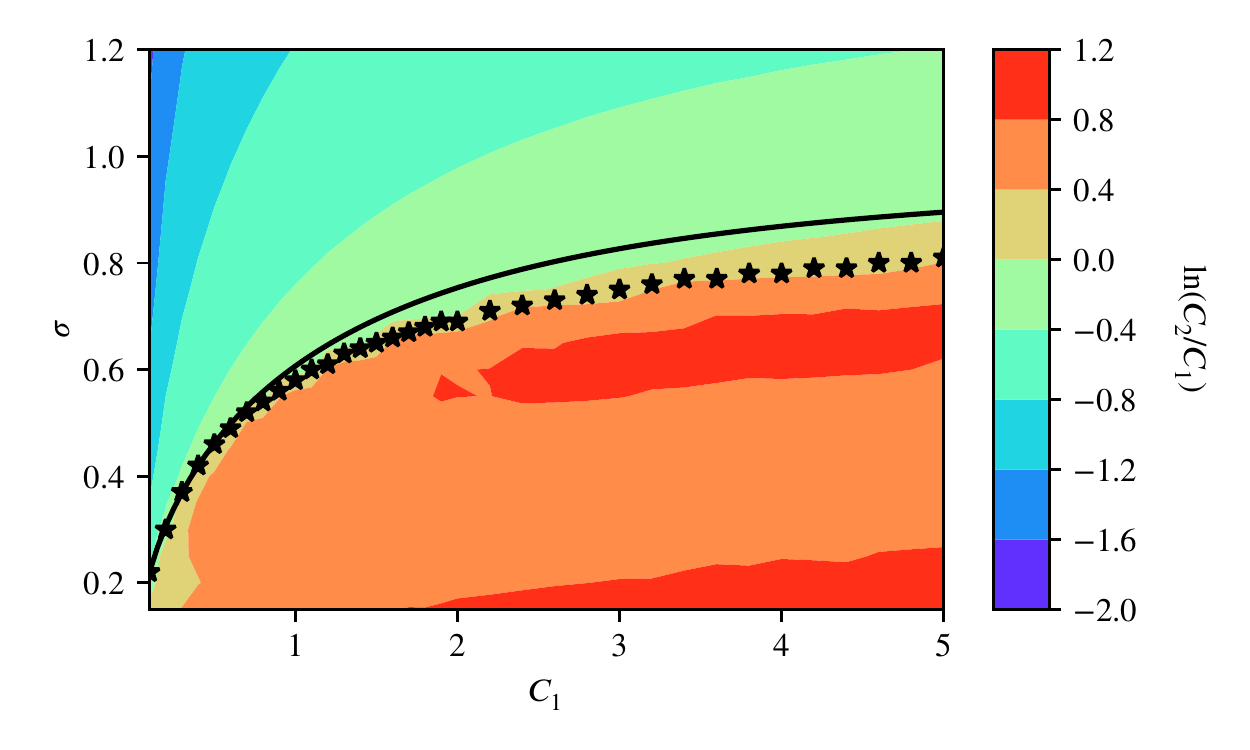}
    \caption{Logarithm of the two-particle correlation parameter $C_2$ divided by the average number of neighbors $C_1$ in dependence on noise strength $\sigma$ and particle density for the model defined in Eq.~\eqref{eq:model}. The black solid line shows the mean field transition towards polar order, whereas the black stars represent the actually measured transition points.
    Simulation parameters: particle number $N=10^5$, velocity $v=1$, interaction radius $R=1$, time step $\Delta t=0.03$, the thermalization time as well as the measurement time was $5\times 10^4$ time units. The results have been averaged over $16$ realizations.
    \label{fig:continuous_ctwo}}
\end{figure}

\begin{figure}[H]
    \centering
    \includegraphics[width=0.9\columnwidth]{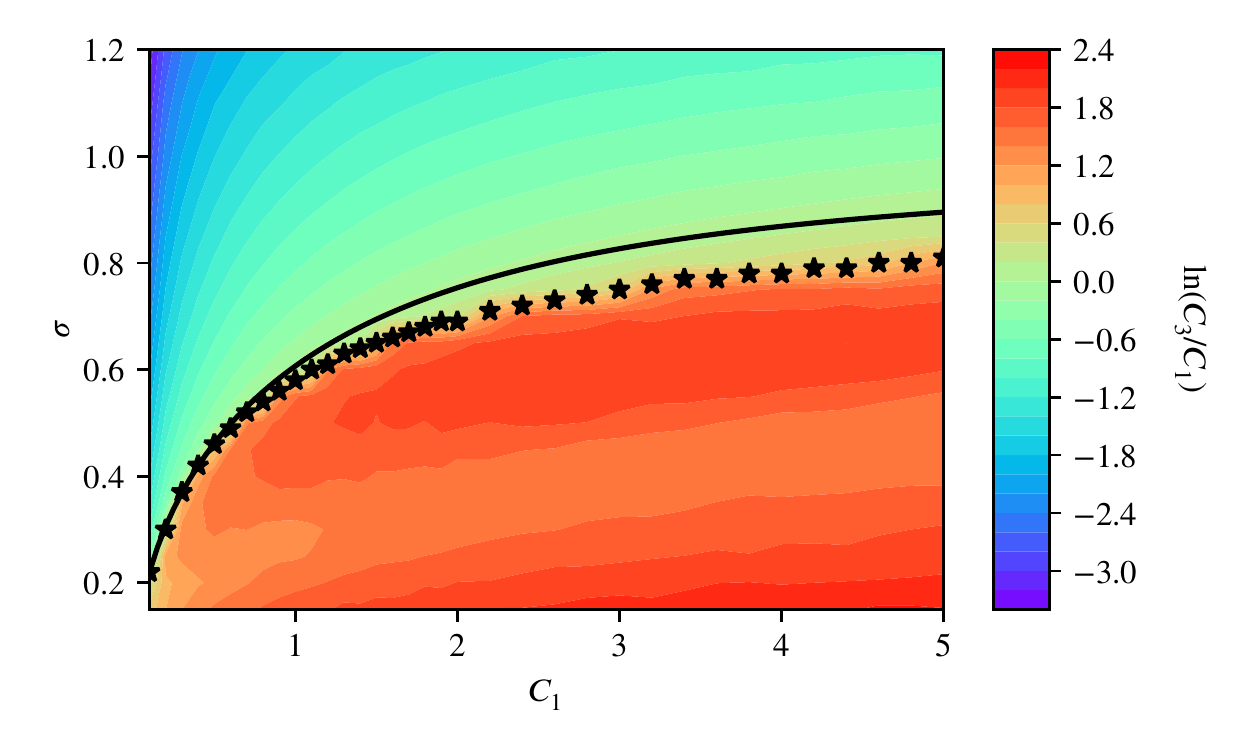}
    \caption{Logarithm of the three-particle correlation parameter $C_3$ divided by the average number of neighbors $C_1$ in dependence on noise strength $\sigma$ and particle density for the model defined in Eq.~\eqref{eq:model}. The black solid line shows the mean field transition towards polar order, whereas the black stars represent the actually measured transition points. Parameters as in Fig.~\ref{fig:continuous_ctwo}.
    \label{fig:continuous_cthree}}
\end{figure}

In Figs.~\ref{fig:continuous_ctwo} and \ref{fig:continuous_cthree} we show the two-particle and three-particle correlation parameters $C_2$ and $C_3$ for the continuous time Vicsek-like model, respectively. The results are qualitatively equivalent to the ones in Fig.~3 of the Letter and Fig.~\ref{fig:cthree}.

\vfill\null
\vspace{5cm}

\begin{thebibliography}{32}%
\makeatletter
\providecommand \@ifxundefined [1]{%
 \@ifx{#1\undefined}
}%
\providecommand \@ifnum [1]{%
 \ifnum #1\expandafter \@firstoftwo
 \else \expandafter \@secondoftwo
 \fi
}%
\providecommand \@ifx [1]{%
 \ifx #1\expandafter \@firstoftwo
 \else \expandafter \@secondoftwo
 \fi
}%
\providecommand \natexlab [1]{#1}%
\providecommand \enquote  [1]{``#1''}%
\providecommand \bibnamefont  [1]{#1}%
\providecommand \bibfnamefont [1]{#1}%
\providecommand \citenamefont [1]{#1}%
\providecommand \href@noop [0]{\@secondoftwo}%
\providecommand \href [0]{\begingroup \@sanitize@url \@href}%
\providecommand \@href[1]{\@@startlink{#1}\@@href}%
\providecommand \@@href[1]{\endgroup#1\@@endlink}%
\providecommand \@sanitize@url [0]{\catcode `\\12\catcode `\$12\catcode
  `\&12\catcode `\#12\catcode `\^12\catcode `\_12\catcode `\%12\relax}%
\providecommand \@@startlink[1]{}%
\providecommand \@@endlink[0]{}%
\providecommand \url  [0]{\begingroup\@sanitize@url \@url }%
\providecommand \@url [1]{\endgroup\@href {#1}{\urlprefix }}%
\providecommand \urlprefix  [0]{URL }%
\providecommand \Eprint [0]{\href }%
\providecommand \doibase [0]{http://dx.doi.org/}%
\providecommand \selectlanguage [0]{\@gobble}%
\providecommand \bibinfo  [0]{\@secondoftwo}%
\providecommand \bibfield  [0]{\@secondoftwo}%
\providecommand \translation [1]{[#1]}%
\providecommand \BibitemOpen [0]{}%
\providecommand \bibitemStop [0]{}%
\providecommand \bibitemNoStop [0]{.\EOS\space}%
\providecommand \EOS [0]{\spacefactor3000\relax}%
\providecommand \BibitemShut  [1]{\csname bibitem#1\endcsname}%
\let\auto@bib@innerbib\@empty
\bibitem [{\citenamefont {Onsager}(1944)}]{Onsager44}%
  \BibitemOpen
  \bibfield  {author} {\bibinfo {author} {\bibfnamefont {L.}~\bibnamefont
  {Onsager}},\ }\href {\doibase 10.1103/PhysRev.65.117} {\bibfield  {journal}
  {\bibinfo  {journal} {Phys. Rev.}\ }\textbf {\bibinfo {volume} {65}},\
  \bibinfo {pages} {117} (\bibinfo {year} {1944})}\BibitemShut {NoStop}%
\bibitem [{\citenamefont {Chou}\ and\ \citenamefont {Ihle}(2015)}]{CI15}%
  \BibitemOpen
  \bibfield  {author} {\bibinfo {author} {\bibfnamefont {Y.-L.}\ \bibnamefont
  {Chou}}\ and\ \bibinfo {author} {\bibfnamefont {T.}~\bibnamefont {Ihle}},\
  }\href@noop {} {\bibfield  {journal} {\bibinfo  {journal} {Phys. Rev. E}\
  }\textbf {\bibinfo {volume} {91}},\ \bibinfo {pages} {022103} (\bibinfo
  {year} {2015})}\BibitemShut {NoStop}%
\bibitem [{\citenamefont {Denton}\ and\ \citenamefont {Ashcroft}(1989)}]{DA89}%
  \BibitemOpen
  \bibfield  {author} {\bibinfo {author} {\bibfnamefont {A.~R.}\ \bibnamefont
  {Denton}}\ and\ \bibinfo {author} {\bibfnamefont {N.~W.}\ \bibnamefont
  {Ashcroft}},\ }\href {\doibase 10.1103/PhysRevA.39.426} {\bibfield  {journal}
  {\bibinfo  {journal} {Phys. Rev. A}\ }\textbf {\bibinfo {volume} {39}},\
  \bibinfo {pages} {426} (\bibinfo {year} {1989})}\BibitemShut {NoStop}%
\bibitem [{\citenamefont {Blawzdziewicz}\ \emph {et~al.}(1989)\citenamefont
  {Blawzdziewicz}, \citenamefont {Cichocki},\ and\ \citenamefont
  {Szamel}}]{BCS89}%
  \BibitemOpen
  \bibfield  {author} {\bibinfo {author} {\bibfnamefont {J.}~\bibnamefont
  {Blawzdziewicz}}, \bibinfo {author} {\bibfnamefont {B.}~\bibnamefont
  {Cichocki}}, \ and\ \bibinfo {author} {\bibfnamefont {G.}~\bibnamefont
  {Szamel}},\ }\href@noop {} {\bibfield  {journal} {\bibinfo  {journal} {J.
  Chem. Phys.}\ }\textbf {\bibinfo {volume} {91}},\ \bibinfo {pages} {7467}
  (\bibinfo {year} {1989})}\BibitemShut {NoStop}%
\bibitem [{\citenamefont {{Mecke}}\ \emph {et~al.}(1994)\citenamefont
  {{Mecke}}, \citenamefont {{Buchert}},\ and\ \citenamefont
  {{Wagner}}}]{MBW94}%
  \BibitemOpen
  \bibfield  {author} {\bibinfo {author} {\bibfnamefont {K.~R.}\ \bibnamefont
  {{Mecke}}}, \bibinfo {author} {\bibfnamefont {T.}~\bibnamefont {{Buchert}}},
  \ and\ \bibinfo {author} {\bibfnamefont {H.}~\bibnamefont {{Wagner}}},\
  }\href@noop {} {\bibfield  {journal} {\bibinfo  {journal} {Astron.
  Astrophys.}\ }\textbf {\bibinfo {volume} {288}},\ \bibinfo {pages} {697}
  (\bibinfo {year} {1994})}\BibitemShut {NoStop}%
\bibitem [{\citenamefont {Diggle}(1983)}]{Diggle83}%
  \BibitemOpen
  \bibfield  {author} {\bibinfo {author} {\bibfnamefont {P.~J.}\ \bibnamefont
  {Diggle}},\ }\href@noop {} {\emph {\bibinfo {title} {Statistical analysis of
  spatial point patterns.}}}\ (\bibinfo  {publisher} {Academic Press},\
  \bibinfo {year} {1983})\BibitemShut {NoStop}%
\bibitem [{\citenamefont {Illian}\ \emph {et~al.}(2008)\citenamefont {Illian},
  \citenamefont {Penttinen}, \citenamefont {Stoyan},\ and\ \citenamefont
  {Stoyan}}]{IPSS08}%
  \BibitemOpen
  \bibfield  {author} {\bibinfo {author} {\bibfnamefont {J.}~\bibnamefont
  {Illian}}, \bibinfo {author} {\bibfnamefont {A.}~\bibnamefont {Penttinen}},
  \bibinfo {author} {\bibfnamefont {H.}~\bibnamefont {Stoyan}}, \ and\ \bibinfo
  {author} {\bibfnamefont {D.}~\bibnamefont {Stoyan}},\ }\href@noop {} {\emph
  {\bibinfo {title} {Statistical analysis and modelling of spatial point
  patterns}}},\ Vol.~\bibinfo {volume} {70}\ (\bibinfo  {publisher} {John Wiley
  \& Sons},\ \bibinfo {year} {2008})\BibitemShut {NoStop}%
\bibitem [{\citenamefont {Baddeley}\ \emph {et~al.}(2015)\citenamefont
  {Baddeley}, \citenamefont {Rubak},\ and\ \citenamefont {Turner}}]{BRT15}%
  \BibitemOpen
  \bibfield  {author} {\bibinfo {author} {\bibfnamefont {A.}~\bibnamefont
  {Baddeley}}, \bibinfo {author} {\bibfnamefont {E.}~\bibnamefont {Rubak}}, \
  and\ \bibinfo {author} {\bibfnamefont {R.}~\bibnamefont {Turner}},\
  }\href@noop {} {\emph {\bibinfo {title} {Spatial point patterns: methodology
  and applications with R}}}\ (\bibinfo  {publisher} {Chapman and Hall/CRC},\
  \bibinfo {year} {2015})\BibitemShut {NoStop}%
\bibitem [{\citenamefont {Law}\ \emph {et~al.}(2009)\citenamefont {Law},
  \citenamefont {Illian}, \citenamefont {Burslem}, \citenamefont {Gratzer},
  \citenamefont {Gunatilleke},\ and\ \citenamefont {Gunatilleke}}]{LIBGGG09}%
  \BibitemOpen
  \bibfield  {author} {\bibinfo {author} {\bibfnamefont {R.}~\bibnamefont
  {Law}}, \bibinfo {author} {\bibfnamefont {J.}~\bibnamefont {Illian}},
  \bibinfo {author} {\bibfnamefont {D.~F.}\ \bibnamefont {Burslem}}, \bibinfo
  {author} {\bibfnamefont {G.}~\bibnamefont {Gratzer}}, \bibinfo {author}
  {\bibfnamefont {C.}~\bibnamefont {Gunatilleke}}, \ and\ \bibinfo {author}
  {\bibfnamefont {I.}~\bibnamefont {Gunatilleke}},\ }\href@noop {} {\bibfield
  {journal} {\bibinfo  {journal} {J. Ecol.}\ }\textbf {\bibinfo {volume}
  {97}},\ \bibinfo {pages} {616} (\bibinfo {year} {2009})}\BibitemShut
  {NoStop}%
\bibitem [{\citenamefont {Vel{\'a}zquez}\ \emph {et~al.}(2016)\citenamefont
  {Vel{\'a}zquez}, \citenamefont {Mart{\'\i}nez}, \citenamefont {Getzin},
  \citenamefont {Moloney},\ and\ \citenamefont {Wiegand}}]{VMGMW16}%
  \BibitemOpen
  \bibfield  {author} {\bibinfo {author} {\bibfnamefont {E.}~\bibnamefont
  {Vel{\'a}zquez}}, \bibinfo {author} {\bibfnamefont {I.}~\bibnamefont
  {Mart{\'\i}nez}}, \bibinfo {author} {\bibfnamefont {S.}~\bibnamefont
  {Getzin}}, \bibinfo {author} {\bibfnamefont {K.~A.}\ \bibnamefont {Moloney}},
  \ and\ \bibinfo {author} {\bibfnamefont {T.}~\bibnamefont {Wiegand}},\
  }\href@noop {} {\bibfield  {journal} {\bibinfo  {journal} {Ecography}\
  }\textbf {\bibinfo {volume} {39}},\ \bibinfo {pages} {1042} (\bibinfo {year}
  {2016})}\BibitemShut {NoStop}%
\bibitem [{\citenamefont {Diggle}\ \emph {et~al.}(1991)\citenamefont {Diggle},
  \citenamefont {Lange},\ and\ \citenamefont {Bene{\v{s}}}}]{DLB91}%
  \BibitemOpen
  \bibfield  {author} {\bibinfo {author} {\bibfnamefont {P.~J.}\ \bibnamefont
  {Diggle}}, \bibinfo {author} {\bibfnamefont {N.}~\bibnamefont {Lange}}, \
  and\ \bibinfo {author} {\bibfnamefont {F.~M.}\ \bibnamefont {Bene{\v{s}}}},\
  }\href@noop {} {\bibfield  {journal} {\bibinfo  {journal} {J. Am. Stat.
  Assoc.}\ }\textbf {\bibinfo {volume} {86}},\ \bibinfo {pages} {618} (\bibinfo
  {year} {1991})}\BibitemShut {NoStop}%
\bibitem [{\citenamefont {Kerscher}\ \emph {et~al.}(1997)\citenamefont
  {Kerscher}, \citenamefont {Schmalzing}, \citenamefont {Retzlaff},
  \citenamefont {Borgani}, \citenamefont {Buchert}, \citenamefont {Gottlöber},
  \citenamefont {Müller}, \citenamefont {Plionis},\ and\ \citenamefont
  {Wagner}}]{KSRBBGMPW97}%
  \BibitemOpen
  \bibfield  {author} {\bibinfo {author} {\bibfnamefont {M.}~\bibnamefont
  {Kerscher}}, \bibinfo {author} {\bibfnamefont {J.}~\bibnamefont
  {Schmalzing}}, \bibinfo {author} {\bibfnamefont {J.}~\bibnamefont
  {Retzlaff}}, \bibinfo {author} {\bibfnamefont {S.}~\bibnamefont {Borgani}},
  \bibinfo {author} {\bibfnamefont {T.}~\bibnamefont {Buchert}}, \bibinfo
  {author} {\bibfnamefont {S.}~\bibnamefont {Gottlöber}}, \bibinfo {author}
  {\bibfnamefont {V.}~\bibnamefont {Müller}}, \bibinfo {author} {\bibfnamefont
  {M.}~\bibnamefont {Plionis}}, \ and\ \bibinfo {author} {\bibfnamefont
  {H.}~\bibnamefont {Wagner}},\ }\href@noop {} {\bibfield  {journal} {\bibinfo
  {journal} {Mon. Not. R. Astron. Soc.}\ }\textbf {\bibinfo {volume} {284}},\
  \bibinfo {pages} {73} (\bibinfo {year} {1997})}\BibitemShut {NoStop}%
\bibitem [{\citenamefont {Parker}\ \emph {et~al.}(2013)\citenamefont {Parker},
  \citenamefont {Sherman}, \citenamefont {van~de Raa}, \citenamefont {van~der
  Meer}, \citenamefont {Samelson},\ and\ \citenamefont {Losert}}]{PSRMSL13}%
  \BibitemOpen
  \bibfield  {author} {\bibinfo {author} {\bibfnamefont {J.}~\bibnamefont
  {Parker}}, \bibinfo {author} {\bibfnamefont {E.}~\bibnamefont {Sherman}},
  \bibinfo {author} {\bibfnamefont {M.}~\bibnamefont {van~de Raa}}, \bibinfo
  {author} {\bibfnamefont {D.}~\bibnamefont {van~der Meer}}, \bibinfo {author}
  {\bibfnamefont {L.~E.}\ \bibnamefont {Samelson}}, \ and\ \bibinfo {author}
  {\bibfnamefont {W.}~\bibnamefont {Losert}},\ }\href@noop {} {\bibfield
  {journal} {\bibinfo  {journal} {Physical Review E}\ }\textbf {\bibinfo
  {volume} {88}},\ \bibinfo {pages} {022720} (\bibinfo {year}
  {2013})}\BibitemShut {NoStop}%
\bibitem [{\citenamefont {Mantz}\ \emph {et~al.}(2008)\citenamefont {Mantz},
  \citenamefont {Jacobs},\ and\ \citenamefont {Mecke}}]{MJM08}%
  \BibitemOpen
  \bibfield  {author} {\bibinfo {author} {\bibfnamefont {H.}~\bibnamefont
  {Mantz}}, \bibinfo {author} {\bibfnamefont {K.}~\bibnamefont {Jacobs}}, \
  and\ \bibinfo {author} {\bibfnamefont {K.}~\bibnamefont {Mecke}},\
  }\href@noop {} {\bibfield  {journal} {\bibinfo  {journal} {J. Stat. Mech. -
  Theory E.}\ }\textbf {\bibinfo {volume} {2008}},\ \bibinfo {pages} {P12015}
  (\bibinfo {year} {2008})}\BibitemShut {NoStop}%
\bibitem [{\citenamefont {Vicsek}\ \emph {et~al.}(1995)\citenamefont {Vicsek},
  \citenamefont {Czir{\'o}k}, \citenamefont {Ben-Jacob}, \citenamefont
  {Cohen},\ and\ \citenamefont {Shochet}}]{VCBCS95}%
  \BibitemOpen
  \bibfield  {author} {\bibinfo {author} {\bibfnamefont {T.}~\bibnamefont
  {Vicsek}}, \bibinfo {author} {\bibfnamefont {A.}~\bibnamefont {Czir{\'o}k}},
  \bibinfo {author} {\bibfnamefont {E.}~\bibnamefont {Ben-Jacob}}, \bibinfo
  {author} {\bibfnamefont {I.}~\bibnamefont {Cohen}}, \ and\ \bibinfo {author}
  {\bibfnamefont {O.}~\bibnamefont {Shochet}},\ }\href@noop {} {\bibfield
  {journal} {\bibinfo  {journal} {Phys. Rev. Lett.}\ }\textbf {\bibinfo
  {volume} {75}},\ \bibinfo {pages} {1226} (\bibinfo {year}
  {1995})}\BibitemShut {NoStop}%
\bibitem [{\citenamefont {Czir{\'o}k}\ \emph {et~al.}(1997)\citenamefont
  {Czir{\'o}k}, \citenamefont {Stanley},\ and\ \citenamefont {Vicsek}}]{CSV97}%
  \BibitemOpen
  \bibfield  {author} {\bibinfo {author} {\bibfnamefont {A.}~\bibnamefont
  {Czir{\'o}k}}, \bibinfo {author} {\bibfnamefont {H.~E.}\ \bibnamefont
  {Stanley}}, \ and\ \bibinfo {author} {\bibfnamefont {T.}~\bibnamefont
  {Vicsek}},\ }\href@noop {} {\bibfield  {journal} {\bibinfo  {journal} {J.
  Phys. A - Math. Gen.}\ }\textbf {\bibinfo {volume} {30}},\ \bibinfo {pages}
  {1375} (\bibinfo {year} {1997})}\BibitemShut {NoStop}%
\bibitem [{\citenamefont {Gr{\'e}goire}\ and\ \citenamefont
  {Chat{\'e}}(2004)}]{GC04}%
  \BibitemOpen
  \bibfield  {author} {\bibinfo {author} {\bibfnamefont {G.}~\bibnamefont
  {Gr{\'e}goire}}\ and\ \bibinfo {author} {\bibfnamefont {H.}~\bibnamefont
  {Chat{\'e}}},\ }\href@noop {} {\bibfield  {journal} {\bibinfo  {journal}
  {Phys. Rev. Lett.}\ }\textbf {\bibinfo {volume} {92}},\ \bibinfo {pages}
  {025702} (\bibinfo {year} {2004})}\BibitemShut {NoStop}%
\bibitem [{\citenamefont {Chat{\'e}}\ \emph {et~al.}(2008)\citenamefont
  {Chat{\'e}}, \citenamefont {Ginelli}, \citenamefont {Gr{\'e}goire},\ and\
  \citenamefont {Raynaud}}]{CGGR08}%
  \BibitemOpen
  \bibfield  {author} {\bibinfo {author} {\bibfnamefont {H.}~\bibnamefont
  {Chat{\'e}}}, \bibinfo {author} {\bibfnamefont {F.}~\bibnamefont {Ginelli}},
  \bibinfo {author} {\bibfnamefont {G.}~\bibnamefont {Gr{\'e}goire}}, \ and\
  \bibinfo {author} {\bibfnamefont {F.}~\bibnamefont {Raynaud}},\ }\href@noop
  {} {\bibfield  {journal} {\bibinfo  {journal} {Phy. Rev. E}\ }\textbf
  {\bibinfo {volume} {77}},\ \bibinfo {pages} {046113} (\bibinfo {year}
  {2008})}\BibitemShut {NoStop}%
\bibitem [{\citenamefont {Solon}\ \emph
  {et~al.}(2015{\natexlab{a}})\citenamefont {Solon}, \citenamefont {Chat\'e},\
  and\ \citenamefont {Tailleur}}]{SCT15}%
  \BibitemOpen
  \bibfield  {author} {\bibinfo {author} {\bibfnamefont {A.~P.}\ \bibnamefont
  {Solon}}, \bibinfo {author} {\bibfnamefont {H.}~\bibnamefont {Chat\'e}}, \
  and\ \bibinfo {author} {\bibfnamefont {J.}~\bibnamefont {Tailleur}},\ }\href
  {\doibase 10.1103/PhysRevLett.114.068101} {\bibfield  {journal} {\bibinfo
  {journal} {Phys. Rev. Lett.}\ }\textbf {\bibinfo {volume} {114}},\ \bibinfo
  {pages} {068101} (\bibinfo {year} {2015}{\natexlab{a}})}\BibitemShut
  {NoStop}%
\bibitem [{\citenamefont {Solon}\ \emph
  {et~al.}(2015{\natexlab{b}})\citenamefont {Solon}, \citenamefont {Caussin},
  \citenamefont {Bartolo}, \citenamefont {Chat\'e},\ and\ \citenamefont
  {Tailleur}}]{SCBCT15}%
  \BibitemOpen
  \bibfield  {author} {\bibinfo {author} {\bibfnamefont {A.~P.}\ \bibnamefont
  {Solon}}, \bibinfo {author} {\bibfnamefont {J.-B.}\ \bibnamefont {Caussin}},
  \bibinfo {author} {\bibfnamefont {D.}~\bibnamefont {Bartolo}}, \bibinfo
  {author} {\bibfnamefont {H.}~\bibnamefont {Chat\'e}}, \ and\ \bibinfo
  {author} {\bibfnamefont {J.}~\bibnamefont {Tailleur}},\ }\href {\doibase
  10.1103/PhysRevE.92.062111} {\bibfield  {journal} {\bibinfo  {journal} {Phys.
  Rev. E}\ }\textbf {\bibinfo {volume} {92}},\ \bibinfo {pages} {062111}
  (\bibinfo {year} {2015}{\natexlab{b}})}\BibitemShut {NoStop}%
\bibitem [{\citenamefont {Poisson}(1837)}]{Poisson1837}%
  \BibitemOpen
  \bibfield  {author} {\bibinfo {author} {\bibfnamefont {S.~D.}\ \bibnamefont
  {Poisson}},\ }\href@noop {} {\emph {\bibinfo {title} {Probabilité des
  jugements en matière criminelle et en matière civile, précédées des
  règles générales du calcul des probabilitiés}}}\ (\bibinfo  {publisher}
  {Bachelier, Paris, France},\ \bibinfo {year} {1837})\BibitemShut {NoStop}%
\bibitem [{Note1()}]{Note1}%
  \BibitemOpen
  \bibinfo {note} {\label {fn:supp}See supplemental material at pages
  7-19}\BibitemShut {NoStop}%
\bibitem [{\citenamefont {Ursell}(1927)}]{Ursell27}%
  \BibitemOpen
  \bibfield  {author} {\bibinfo {author} {\bibfnamefont {H.~D.}\ \bibnamefont
  {Ursell}},\ }\href {\doibase 10.1017/S0305004100011191} {\bibfield  {journal}
  {\bibinfo  {journal} {Math. Proc. Cambridge}\ }\textbf {\bibinfo {volume}
  {23}},\ \bibinfo {pages} {685–697} (\bibinfo {year} {1927})}\BibitemShut
  {NoStop}%
\bibitem [{\citenamefont {Mayer}\ and\ \citenamefont {Montroll}(1941)}]{MM41}%
  \BibitemOpen
  \bibfield  {author} {\bibinfo {author} {\bibfnamefont {J.~E.}\ \bibnamefont
  {Mayer}}\ and\ \bibinfo {author} {\bibfnamefont {E.}~\bibnamefont
  {Montroll}},\ }\href@noop {} {\bibfield  {journal} {\bibinfo  {journal} {J.
  Chem. Phys.}\ }\textbf {\bibinfo {volume} {9}},\ \bibinfo {pages} {2}
  (\bibinfo {year} {1941})}\BibitemShut {NoStop}%
\bibitem [{Note2()}]{Note2}%
  \BibitemOpen
  \bibinfo {note} {The overlap area of three arbitrarily placed circles
  generalizes the area formula of the famous Reuleaux triangle \cite
  {Reuleaux1875}.}\BibitemShut {Stop}%
\bibitem [{\citenamefont {Bricard}\ \emph {et~al.}(2013)\citenamefont
  {Bricard}, \citenamefont {Caussin}, \citenamefont {Desreumaux}, \citenamefont
  {Dauchot},\ and\ \citenamefont {Bartolo}}]{BCDDB13}%
  \BibitemOpen
  \bibfield  {author} {\bibinfo {author} {\bibfnamefont {A.}~\bibnamefont
  {Bricard}}, \bibinfo {author} {\bibfnamefont {J.-B.}\ \bibnamefont
  {Caussin}}, \bibinfo {author} {\bibfnamefont {N.}~\bibnamefont {Desreumaux}},
  \bibinfo {author} {\bibfnamefont {O.}~\bibnamefont {Dauchot}}, \ and\
  \bibinfo {author} {\bibfnamefont {D.}~\bibnamefont {Bartolo}},\ }\href@noop
  {} {\bibfield  {journal} {\bibinfo  {journal} {Nature}\ }\textbf {\bibinfo
  {volume} {503}},\ \bibinfo {pages} {95} (\bibinfo {year} {2013})}\BibitemShut
  {NoStop}%
\bibitem [{\citenamefont {Kullback}\ and\ \citenamefont
  {Leibler}(1951)}]{KL51}%
  \BibitemOpen
  \bibfield  {author} {\bibinfo {author} {\bibfnamefont {S.}~\bibnamefont
  {Kullback}}\ and\ \bibinfo {author} {\bibfnamefont {R.~A.}\ \bibnamefont
  {Leibler}},\ }\href {\doibase 10.1214/aoms/1177729694} {\bibfield  {journal}
  {\bibinfo  {journal} {Ann. Math. Statist.}\ }\textbf {\bibinfo {volume}
  {22}},\ \bibinfo {pages} {79} (\bibinfo {year} {1951})}\BibitemShut {NoStop}%
\bibitem [{\citenamefont {Toner}\ and\ \citenamefont {Tu}(1995)}]{TT95}%
  \BibitemOpen
  \bibfield  {author} {\bibinfo {author} {\bibfnamefont {J.}~\bibnamefont
  {Toner}}\ and\ \bibinfo {author} {\bibfnamefont {Y.}~\bibnamefont {Tu}},\
  }\href@noop {} {\bibfield  {journal} {\bibinfo  {journal} {Phys. Rev. Lett.}\
  }\textbf {\bibinfo {volume} {75}},\ \bibinfo {pages} {4326} (\bibinfo {year}
  {1995})}\BibitemShut {NoStop}%
\bibitem [{\citenamefont {Toner}\ and\ \citenamefont {Tu}(1998)}]{TT98}%
  \BibitemOpen
  \bibfield  {author} {\bibinfo {author} {\bibfnamefont {J.}~\bibnamefont
  {Toner}}\ and\ \bibinfo {author} {\bibfnamefont {Y.}~\bibnamefont {Tu}},\
  }\href@noop {} {\bibfield  {journal} {\bibinfo  {journal} {Phys. Rev. E}\
  }\textbf {\bibinfo {volume} {58}},\ \bibinfo {pages} {4828} (\bibinfo {year}
  {1998})}\BibitemShut {NoStop}%
\bibitem [{\citenamefont {Ihle}(2013)}]{Ihle13}%
  \BibitemOpen
  \bibfield  {author} {\bibinfo {author} {\bibfnamefont {T.}~\bibnamefont
  {Ihle}},\ }\href@noop {} {\bibfield  {journal} {\bibinfo  {journal} {Phys.
  Rev. E}\ }\textbf {\bibinfo {volume} {88}},\ \bibinfo {pages} {040303}
  (\bibinfo {year} {2013})}\BibitemShut {NoStop}%
\bibitem [{Note3()}]{Note3}%
  \BibitemOpen
  \bibinfo {note} {In \cite {SCT15} the density was varied at constant noise
  strength. Therefore both transitions occur not exactly at the same density
  $\rho =0.25$ that was used in Fig.~\ref {fig:transitions}. The polar ordering
  transition occurred at $\rho =0.251$ for $\eta =0.3$ in Ref.~\cite {SCT15}
  from which we interpolated together with the next point $\eta _c=0.29944$ for
  $\rho =0.25$.}\BibitemShut {Stop}%
\bibitem [{\citenamefont {Reuleaux}(1875)}]{Reuleaux1875}%
  \BibitemOpen
  \bibfield  {author} {\bibinfo {author} {\bibfnamefont {F.}~\bibnamefont
  {Reuleaux}},\ }\href@noop {} {\emph {\bibinfo {title} {Lehrbuch der
  Kinematik: Theoretische Kinematik: Grundz{\"u}ge einer Theorie des
  Maschinenwesens. 1}}},\ Vol.~\bibinfo {volume} {1}\ (\bibinfo  {publisher}
  {Vieweg},\ \bibinfo {year} {1875})\BibitemShut {NoStop}%
\end{thebibliography}
\end{document}